\pdfoutput=1
\documentclass[aps,prl,10pt,twocolumn,groupedaddress,amsmath,amssymb,nofootinbib,longbibliography]{revtex4-2}
\usepackage[T1]{fontenc}
\usepackage[utf8]{inputenc}
\usepackage{lmodern}
\usepackage{mathtools,amsthm}
\usepackage{graphicx}
\usepackage{xcolor}
\usepackage[hidelinks]{hyperref}
\usepackage{xurl}
\usepackage{microtype}
\usepackage{todonotes}
\usepackage{soul,perpage,marginnote}
\allowdisplaybreaks
\newtheorem{theorem}{Theorem}
\newtheorem{lemma}[theorem]{Lemma}

\newcommand{\Tr}{\operatorname{Tr}}

\newcommand{\one}{\mathbf 1}

\newcommand{\ii}{\mathrm{i}}
\newcommand{\SU}{\operatorname{SU}}
\newcommand{\SO}{\operatorname{SO}}
\newcommand{\Spin}{\operatorname{Spin}}
\newcommand{\HH}{\mathcal H}
\newcommand{\C}{\mathbb C}
\newcommand{\braket}[2]{\langle #1|#2\rangle}
\newcommand{\me}[3]{\langle #1|#2|#3\rangle}
\newcommand{\norm}[1]{\lVert #1\rVert}
\newcommand{\abs}[1]{\lvert #1\rvert}
\newcommand{\tr}{\operatorname{tr}}
\definecolor{claimnote1}{HTML}{FFF0B3}
\definecolor{claimnote2}{HTML}{D6EAF8}
\definecolor{claimnote3}{HTML}{D9EED3}
\definecolor{claimnote4}{HTML}{EADCF4}
\newcounter{claimnote}
\MakePerPage{claimnote}
\soulregister\cite7
\ExplSyntaxOn
\tl_new:N \l_claimsupport_color_tl
\tl_new:N \l_claimsupport_text_tl
\cs_new_protected:Npn \prl_todo_start:n #1
  {
    \leavevmode
    \stepcounter{claimnote}
    \tl_set:Nx \l_claimsupport_color_tl
      { claimnote\int_eval:n { 1 + \int_mod:nn { \value{claimnote} - 1 } {4} } }
    \colorlet{prlTodoCurrent}{\l_claimsupport_color_tl}
    \todo[noline,nolist,size=\tiny,backgroundcolor=prlTodoCurrent,
      bordercolor=prlTodoCurrent!65!black]{\raggedright #1}
    \sethlcolor{prlTodoCurrent}
  }
\NewDocumentCommand{\claimsupport}{s m m}
  {
    \group_begin:
    \IfBooleanT{#1}{\cs_set_eq:NN \marginpar \marginnote}
    \prl_todo_start:n {#3}
    \tl_set:Nn \l_claimsupport_text_tl {#2}
    \tl_replace_all:Nnn \l_claimsupport_text_tl {\(} {$}
    \tl_replace_all:Nnn \l_claimsupport_text_tl {\)} {$}
    \exp_args:NV \hl \l_claimsupport_text_tl
    \group_end:
  }
\NewDocumentCommand{\prlcolumnmarginpar}{o m}
  {
    \group_begin:
    \cs_set_eq:cN {if@twocolumn} \iftrue
    \IfNoValueTF{#1}
      {\marginnote{#2}}
      {\marginnote[{#1}]{#2}}
    \group_end:
  }
\NewDocumentEnvironment{condensepassage}{}
  {
    \group_begin:
    \cs_set_eq:NN \marginpar \prlcolumnmarginpar
    \prl_todo_start:n {Condense.}
  }
  {\group_end:}
\ExplSyntaxOff

\newcommand{\ket}[1]{\lvert #1\rangle}
\newcommand{\bra}[1]{\langle #1\rvert}

\usepackage[normalem]{ulem}
\newcommand{\authorstrike}[1]{%
  \bgroup\color{black}%
  \markoverwith{\textcolor{#1}{\rule[.5ex]{2pt}{.4pt}}}%
  \ULon}

\newcommand{\figA}{0.94}\newcommand{\figB}{0.94}\newcommand{\figC}{0.46}
\graphicspath{{figures/}}
\usepackage{reported-results}

\begin{document}

\title{\boldmath $99.9\text{...}\!\%$ of the \(\SU(2)\) BFSS ground state}

\author{Alexander Michel}
\email{amichel@fas.harvard.edu}
\author{Jaroslav Scheinpflug}
\email{jscheinpflug@g.harvard.edu}
\author{Xi Yin}
\email{xiyin@fas.harvard.edu}
\affiliation{Jefferson Physical Laboratory, Harvard University, Cambridge, MA, USA}

\begin{abstract}
We solve for the ground-state wavefunction of the \(\SU(2)\) BFSS matrix
model by applying the Rayleigh--Ritz method to an auxiliary operator
\(H_{\mathrm{aux}}\). We show that this operator agrees with the Hamiltonian
on the ground-state subspace and has a spectral gap.
For \(\SU(2)\) specifically, the representation theory of the singlet sector gives an astronomical reduction in the number of degrees of freedom.
This allows us to construct an efficient variational basis whose inner products we compute exactly.
We demonstrate the method by computing a state that
satisfies rigorous bounds on its energy
\(E\,g_{\mathrm{YM}}^{-2/3}\leq\ReportedEnergy{}\)
and ground-subspace overlap
\(\|P_{\ker H}\psi\|^2\geq\ReportedFidelity{}\).
We compute ground-state moments with high precision,
finding values consistent with the bounds from our matrix bootstrap.
Additionally, the state accurately reproduces both the exponent and coefficient of the asymptotic
radial probability density predicted by the three-graviton
amplitude.
\end{abstract}

\maketitle

\section{Introduction}
\label{sec:introduction}

The $U(N)$ Banks--Fischler--Shenker--Susskind matrix quantum mechanics \cite{BFSS,Polchinski99,MaldacenaStrings24,LinTASI} is the unique interacting quantum mechanics with maximal supersymmetry and $\Spin(9)$ R-symmetry and can be obtained from a dimensional reduction of four-dimensional $U(N)$ $\mathcal{N}=4$ super Yang--Mills theory, with degrees of freedom described by a $\Spin(9)$ vector $X^I$ and a $\Spin(9)$ spinor $\theta^\alpha$, both in the adjoint of $U(N)$. It arises as the low-energy description of open strings on a stack of $N$ D0-branes in type IIA string theory in flat ten-dimensional spacetime. As such, it plays a key role in the IIA/M-theory duality, with the M-theory supergraviton carrying $N$ units of Kaluza--Klein momentum along the M-theory circle arising from a threshold bound state of $N$ D0-branes. From the D0-brane picture and the IIA/M-theory duality, the model itself is holographically dual~\cite{Maldacena97,BGL,IMSY} to M-theory on a compactified pp-wave background carrying $N$ units of momentum along the compactified direction, with a singular, asymptotically lightlike compactification. The threshold bound state is dual to the M-theory vacuum geometry itself.

Finding the threshold bound state is a problem of constructing the gauge-invariant $\Spin(9)$-singlet ground state of the $\SU(N)$ BFSS that remains after decoupling the free $U(1)$ subsector of the $U(N)$ BFSS. Proving the uniqueness of this ground state is notoriously subtle because the $\SU(N)$ BFSS has no gap. For $\SU(2)$, a first-principle computation of the Witten index~\cite{SethiStern,Yi97} shows its existence, and combined with the $\Spin(9)$ invariance of every zero-energy state~\cite{SethiStern00,HaslerHoppe} also its uniqueness. For general $N$, the index was computed under additional assumptions~\cite{PorratiRozenberg,GreenGutperle,MNS,KacSmilga,Staudacher,LeeYi}, and a proof of uniqueness was recently proposed~\cite{OpenAIBFSS}.

Beyond existence and uniqueness, it was widely believed that constructing the ground state globally was out of reach of conventional methods such as Rayleigh--Ritz (though see results on its local form~\cite{HalpernSchwartz,Konechny,GrafHoppe,FroehlichEtAl,HaslerHoppeAsym,HoppeLundholmTrzetrzelewski,MichishitaTrzetrzelewski,LinYin}), as even in the simplest case of $\SU(2)$, the ground state consists of $2^{24}$ components, each a function of $27$ real variables. In this work, we leverage the surprising fact that, for $\SU(2)$ specifically, representation theory enables an astronomical reduction in the degrees of freedom, allowing us to specify any ground-state configuration with only $2066$ components, each a function of $3$ real variables. As we explain, for $\SU(N)$ with $N \geq 3$, representation theory offers little to no simplification.

The reduction makes a systematic variational treatment possible:
a Hermite basis in each of the $2066$ components already spans a dense
subspace of the singlet sector. Our finite trial space improves on this idea in two key ways. We
first augment the isotropic spatial Gaussian $\exp(-\sum_i\tr\mathbf{X}^2_i)$ with a profile that
extends into the commuting valley, motivated by known power-law decay there~\cite{HalpernSchwartz,GrafHoppe,FroehlichEtAl}. Then we generate our trial basis from a small number of polynomial singlets by acting with powers of the Yukawa coupling and scalar invariants. The Yukawas probe additional fermionic structures, and the scalar invariants further increase polynomial order. By construction, all expectations and overlaps of interest can be computed exactly using Wick contractions and elementary integrals.

Even after this reduction, an obstruction remains to the standard application
of the Rayleigh--Ritz method: the continuum reaches zero, so a small
variational energy alone cannot establish that a trial state is close to the
ground state. We find a simple auxiliary Hamiltonian,
$H_{\rm aux}=\tfrac1{16}\sum_\alpha Q_\alpha(\sum_i\tr\mathbf X_i^2)Q_\alpha$,
where $Q_\alpha$ are the sixteen supercharges and $\mathbf X_i$ are the nine
bosonic matrices. It has the same
zero-energy ground subspace as the BFSS Hamiltonian, and its matrix elements in
our trial space can be evaluated analytically. Using R-symmetry, we prove a
purely kinematic technical lemma that establishes a positive spectral gap for
$H_{\rm aux}$ in the gauge-invariant $\Spin(9)$-singlet sector. This lemma
can be rephrased as the positive semidefiniteness of a
matrix of expectation values. As in the matrix
bootstrap~\cite{Lin20,HanHartnollKruthoff,KazakovZheng,LinD0Bootstrap}, the
bound can then be tightened with semidefinite programming by including more operators
and the dynamics.

A small expectation value of $H_\mathrm{aux}$ therefore guarantees a large overlap
with the BFSS ground space, without assuming uniqueness. Minimizing this
expectation value, we obtain a gauge-invariant $\Spin(9)$-singlet
wavefunction with
$E\leq\ReportedEnergy{}$ in units of $g_{\mathrm{YM}}^{2/3}$ and
ground-space fidelity $F\geq\ReportedFidelity{}$.

We test the spatial profile using the probability density $p(r)$ of the
largest singular value $r$, which measures separation along the valley.
Our state agrees closely with the predicted asymptotic exponent and
coefficient of the radial probability density $p(r)\sim1890\sqrt2\,r^{-10}$. The coefficient prediction follows from the
three-point amplitude of Herderschee and Maldacena~\cite{HerderscheeMaldacena} as we derive in Subsection~\ref{sec:amplitude} and Appendix~\ref{app:vertex}, and the decay power has long been known from the asymptotic analysis of \cite{GrafHoppe,FroehlichEtAl,LinYin}.
Beyond the tail, the state has $\langle R^2\rangle=\ReportedRadiusSquared{}$ and $\langle R^4\rangle=\ReportedRadiusFourth{}$, with $R^2=\sum_{a,i}X_{ai}^2$. These values lie inside the regions allowed by a matrix bootstrap~\cite{Lin20,HanHartnollKruthoff,KazakovZheng,LinD0Bootstrap,LinTASI}, which imposes positivity together with the Ward identities that follow from the ground state being annihilated by the Hamiltonian, the supercharges, the Gauss law and the rotations; at level 10 it certifies $\langle R^2\rangle\geq7.798$ (Subsection~\ref{sec:wavefunction}). Our bootstrap bounds are comparatively loose, so this agreement is a consistency check rather than a precision test.

Section~\ref{sec:model} introduces the model and the singlet
condition obeyed by its zero-energy states.
Section~\ref{sec:symmetry-reduction} uses it to reduce the singlet sector, a
reduction we derive in Appendix~\ref{sec:fiber-proof}, and builds our
variational method on it. The fidelity bound that justifies the variational
technique is proved in Appendix~\ref{app:fidelity} and the matrix elements
needed by it are computed exactly in Appendix~\ref{app:ansatz-wick}.
Section~\ref{sec:results} presents our results. We show that our state
reproduces the predicted asymptotic decay of the radial probability density, whose
power follows from the Coulomb-branch EFT and whose coefficient we derive from the
three-point amplitude in Appendix~\ref{app:vertex}. We then report the energy and fidelity bounds and the moments
computed for our state, and compare the moments with our bootstrap.
We summarize and discuss our results in Section~\ref{sec:discussion}. Our
code~\cite{CompanionCode} implements the whole construction, from the basis
to the certified observables, contains the coefficients of the state in the
fermion conventions we fix in Appendix~\ref{sec:fermions}, and computes our
bootstrap bounds.

\section{The model and its Hilbert space}
\label{sec:model}

Removing the free center of mass of two D0-branes leaves the
interacting $\SU(2)$ model. Color indices $a,b,c=1,2,3$ label the adjoint
of $\SU(2)$, and $i,j,k=1,\ldots,9$ and $\alpha,\beta=1,\ldots,16$ the
vector and spinor of $\Spin(9)$. The bosonic matrices are
\hypertarget{def:X}{}$\mathbf X_i=\sum_aX_{ai}T_a$, with $T_a=\sigma_a/\sqrt2$ and
$\tr(T_aT_b)=\delta_{ab}$; we collect them into
$X=(X_{ai})\in\mathbb R^{3\times9}$, with rows $X_a$.
The fermionic matrices are
$\boldsymbol\theta_\alpha=\sum_a\theta^a_\alpha T_a$, where the 48
Majorana operators obey
$\{\theta^a_\alpha,\theta^b_\beta\}=\delta_{ab}\delta_{\alpha\beta}$.
We use real symmetric $16\times16$ matrices $\Gamma_i$ satisfying
$\{\Gamma_i,\Gamma_j\}=2\delta_{ij}\one$, with the explicit representation
given in Appendix~\ref{sec:fermions}.

In units where $g_{\mathrm{YM}}=1$, the Hamiltonian
is~\cite{LinYin}
\begin{equation}
 \begin{aligned}
 H={}&-\frac12\Delta_X
       -\frac14\sum_{i,j}\tr[\mathbf X_i,\mathbf X_j]^2\\
    &-\frac12\sum_{i,\alpha,\beta}(\Gamma_i)_{\alpha\beta}
       \tr\!\left(\boldsymbol\theta_\alpha
          [\mathbf X_i,\boldsymbol\theta_\beta]\right),
 \end{aligned}
 \label{eq:model-H}
\end{equation}
where $\Delta_X=\sum_{a,i}\partial_{X_{ai}}^2$.
We define the invariant coordinates
\begin{equation}
 \begin{aligned}
  s&=\tr(XX^{\mathsf T})=x_1^2+x_2^2+x_3^2=R^2,\\
  u&=\frac12\bigl[(\tr XX^{\mathsf T})^2-\tr((XX^{\mathsf T})^2)\bigr]\\
   &=x_1^2x_2^2+x_2^2x_3^2+x_3^2x_1^2,\\
  v&=\det(XX^{\mathsf T})=x_1^2x_2^2x_3^2,
 \end{aligned}
 \label{eq:invariant-coordinates}
\end{equation}
where $x_1\geq x_2\geq x_3\geq0$ are the singular values of $X$.
The bosonic potential is
$u=-\frac14\sum_{i,j}\tr[\mathbf X_i,\mathbf X_j]^2$.
It vanishes exactly when all nine $\mathbf X_i$ commute,
equivalently when $\operatorname{rank}X\leq1$.
Writing the fermionic Yukawa interaction as $Y(X)$, the Hamiltonian takes the
component form (with $\epsilon_{123}=1$)
\begin{equation}
 \begin{gathered}
 H=-\frac12\Delta_X+u+Y(X),\\
 Y(X)=\sum_{a,i}X_{ai}K_{ai},\\
 K_{ai}=\frac{\ii}{\sqrt2}\sum_{b,c,\alpha,\beta}\epsilon_{abc}
   (\Gamma_i)_{\alpha\beta}\theta^b_\alpha\theta^c_\beta.
 \end{gathered}
 \label{eq:model-yukawa}
\end{equation}

Pairing the 48 Majorana fermions gives 24 complex modes,
whose Fock space is $\mathcal F\simeq\C^{2^{24}}$. A state
$\ket{\psi}\in L^2(\mathbb R^{27})\otimes\mathcal F$ has wavefunction
$\ket{\psi(X)}=(\bra X\otimes\one)\ket\psi$.
The physical Hilbert space $\HH$ consists of gauge-invariant states,
which satisfy the Gauss law
$(\mathcal L_{ab}-\mathcal G_{ab})\ket\psi=0$, where
\begin{equation}
 \begin{aligned}
 \mathcal L_{ab}&=\sum_i(X_{bi}\partial_{X_{ai}}-X_{ai}\partial_{X_{bi}}),\\
 \mathcal G_{ab}&=\sum_\alpha\theta^a_\alpha\theta^b_\alpha,
 \qquad a\neq b.
 \end{aligned}
 \label{eq:gauge-generators}
\end{equation}

The supercharges, with
$C^\alpha_{ai}=\sum_\beta(\Gamma_i)_{\alpha\beta}\theta^a_\beta$,
\begin{equation}
 \begin{gathered}
 Q_\alpha=-\ii\sum_{a,i}C^\alpha_{ai}\partial_{X_{ai}}+V_\alpha,\\
 V_\alpha=\sqrt2\sum_{i<j,a,b,c,\beta}
 \epsilon_{abc}X_{bi}X_{cj}(\Gamma_i\Gamma_j)_{\alpha\beta}\theta^a_\beta,
 \end{gathered}
 \label{eq:supercharge}
\end{equation}
are gauge invariant, transform as a $\Spin(9)$ spinor and satisfy
\begin{equation}
 \begin{gathered}
 Q_\alpha^2=H+\ii\sqrt2\sum_{a,i}(\Gamma_i)_{\alpha\alpha}X_{ai}
   (\mathcal L_a-\mathcal G_a),\\
 \mathcal G_a=\frac12\sum_{b,c}\epsilon_{abc}\mathcal G_{bc},\quad
 \mathcal L_a=\frac12\sum_{b,c}\epsilon_{abc}\mathcal L_{bc}.
 \end{gathered}
 \label{eq:supercharge-closure}
\end{equation}

Here $\alpha$ is fixed, with no sum. On $\HH$, the Gauss-law
term vanishes, giving $Q_\alpha^2=H$ and
$\me{\psi}{H}{\psi}=\norm{Q_\alpha\psi}^2$.
Thus a physical state has zero energy precisely when all sixteen
supercharges annihilate it.

The $\SU(2)$ gauge symmetry and the $\Spin(9)$ R-symmetry act
by the unitary operators
\begin{equation}
 \begin{gathered}
 \mathcal T(O,V)=\mathcal W(O,V)\otimes\mathcal U(O,V),\\
 \mathcal W(O,V)\ket{X}=\ket{OXV^{\mathsf T}},\\
 O\in\SO(3),\ V\in\SO(9).
 \end{gathered}
 \label{eq:symmetry-operators}
\end{equation}
For rotations in the $ab$ and $ij$ planes, generated by
$(E_{ab})_{cd}=\delta_{ac}\delta_{bd}-\delta_{ad}\delta_{bc}$ and likewise
$E_{ij}$, the fermionic factor is
$\mathcal U(e^{tE_{ab}},1)=e^{t\mathcal G_{ab}}$ and
$\mathcal U(1,e^{tE_{ij}})=e^{t\mathcal R_{ij}}$, with
\begin{equation}
 \begin{gathered}
 \mathcal R_{ij}=\frac14\sum_{a,\alpha,\beta}(\Gamma_i\Gamma_j)_{\alpha\beta}
   \theta^a_\alpha\theta^a_\beta,\qquad i\neq j.
 \end{gathered}
 \label{eq:fermion-rotations}
\end{equation}
Thus $\mathcal T(e^{tE_{ab}},1)=e^{t(\mathcal G_{ab}-\mathcal L_{ab})}$ and
$\mathcal T(1,e^{tE_{ij}})=e^{t(\mathcal R_{ij}+\mathcal S_{ij})}$, with the
bosonic rotation generator
$\mathcal S_{ij}=\sum_a(X_{ai}\partial_{X_{aj}}-X_{aj}\partial_{X_{ai}})$.
Since $K_{ai}$ is a vector of both groups,
$\mathcal U(O,V)Y(X)\mathcal U(O,V)^{-1}=Y(OXV^{\mathsf T})$, and the
$\mathcal T(O,V)$ commute with $H$.

Zero-energy states are moreover
$\Spin(9)$ singlets~\cite{SethiStern00,HaslerHoppe}, and we restrict to
singlets, $\mathcal T(O,V)\ket{\psi}=\ket{\psi}$. Since
$(\bra{OXV^{\mathsf T}}\otimes\one)\mathcal T(O,V)
=\mathcal U(O,V)(\bra{X}\otimes\one)$, their wavefunctions obey the singlet
condition
\begin{equation}
 \ket{\psi(OXV^{\mathsf T})}=\mathcal U(O,V)\ket{\psi(X)}.
 \label{eq:equivariance}
\end{equation}

\section{Symmetry reduction and variational ansatz}
\label{sec:symmetry-reduction}\label{sec:symmetry}

The three rows $X_a$ of $X$ transform as $\SO(9)$ vectors, so we can rotate
them into the first three spatial directions. Using the gauge $\SU(2)\simeq
\SO(3)$ on the columns we can orthogonalize them, leaving only the diagonal
entries $X_{aa}$ nonzero. There remains a large residual symmetry that
fixes the diagonal configuration: the $\SO(6)$ acting on the empty columns
and the simultaneous discrete $\pi$ rotations of the leading three axes. The
fermionic factor must be a singlet under the residual symmetry, which greatly
reduces the number of independent structures that can appear.

For $\SU(N)$ with $N>2$, no comparable reduction occurs. This is a simple
consequence of the fact that $X$ then has more degrees of freedom, $9(N^2-1)$,
than the dimension $N^2+35$ of $\SU(N)\times\Spin(9)$. Hence the stabilizer of a
generic $X$ is just the identity, apart from the usual $-1$ acting on the
fermions, so the singlet condition merely halves the fermionic
components.

\subsection{Structure of the singlet space}
\label{sec:fiber}

For a generic configuration, write
\begin{equation}
 \begin{gathered}
 X=O D(x)V^{\mathsf T},\qquad D(x)_{ai}=x_a\delta_{ai},\\
 x_1>x_2>x_3>0,
 \end{gathered}
\end{equation}
where $O\in\SO(3)$, $V\in\SO(9)$, and $x_a^2$ are the eigenvalues of
$XX^{\mathsf T}$. Configurations with repeated or zero singular values
have measure zero in $\mathbb R^{27}$, so the wavefunction on the generic
set determines the $L^2$ state completely.
The singlet condition~\eqref{eq:equivariance} gives
\begin{equation}
 \ket{\psi(X)}=\mathcal U(O,V)\ket{\psi(D(x))}.
 \label{eq:diagonal-wavefunction}
\end{equation}
Thus the full wavefunction is determined by $\ket{\psi(D(x))}$, which
depends on only three real variables. The remaining symmetry conditions
come from rotations that leave $D(x)$ unchanged.

The stabilizer for all such configurations is the
same,
\begin{equation}
 K_4\times\SO(6),
 \label{eq:diagonal-stabilizer}
\end{equation}
where $K_4\simeq\mathbb Z_2\times\mathbb Z_2$ consists of the identity and
the three distinct rotations by $\pi$ about the principal axes,
acting simultaneously in color and space, and $\SO(6)$ rotates the
last six columns.

The singlet condition requires
$\ket{\psi(D(x))}\in\mathcal F_*$, where
\begin{equation}
 \mathcal F_*=\{\ket\xi\in\mathcal F:
       \mathcal U(g)\ket\xi=\ket\xi\ \text{for every }
       g\in K_4\times\SO(6)\}.
 \label{eq:fixed-fermion-space}
\end{equation}
Using the Schur--Weyl
duality~\cite[Thm.~6.3 and Ex.~6.11(b)]{FultonHarris}, we construct an
explicit basis for $\mathcal F_*$ and show that
\begin{equation}
 \dim\mathcal F_*=2066.
 \label{eq:fiber-dimension}
\end{equation}
We refer the reader to Appendix~\ref{sec:fiber-proof} for the detailed
proof.

Consequently, expanding in the basis~\eqref{eq:fiber-basis-enumerated} of $\mathcal F_*$
gives
\begin{equation}
 \ket{\psi(D(x))}
 =\sum_{\mu=1}^{2066}
   \varphi_\mu(x_1,x_2,x_3)\ket{e_\mu},
 \label{eq:diagonal-components}
\end{equation}
with each $\varphi_\mu$ a scalar function of the singular values in $L^2$.
Equivalently, we can write $\varphi_\mu(x_1,x_2,x_3)=\varphi_\mu(s,u,v)$
as a function of the invariant coordinates defined in
\eqref{eq:invariant-coordinates}.

\subsection{Variational ansatz}
\label{sec:ansatz}
\label{sec:rayleigh-ritz}
\label{sec:profiles}
\label{sec:normalization}
\label{sec:fidelity}

The reduction in~\eqref{eq:diagonal-components} gives a dense family of
singlets. For any fixed $a>0$, the states $\ket{g,\mu}$ with wavefunctions
\begin{equation}
 \begin{gathered}
  \braket{X}{g,\mu}
   \equiv g(s,u,v)e^{-as/2}\mathcal U(O,V)\ket{e_\mu},\\
  X=OD(x)V^{\mathsf T},\qquad \mu=1,\ldots,2066,
 \end{gathered}
 \label{eq:dense-singlet-family}
\end{equation}
span a dense subspace of the singlet sector as $g$ ranges over polynomials
in $s,u,v$.

We now construct a finite variational basis. To do so, we first
define singlets that are polynomials in~$X$,
\begin{equation}
 \ket{P_j(X)}=s^{k_j}u^{l_j}v^{m_j}Y^{p_j}\ket{\Lambda_{\sigma_j}(X)},
 \label{eq:polynomial-factors}
\end{equation}
where the seeds $\ket{\Lambda_\sigma(X)}$ are themselves polynomials in~$X$.
For the variational state reported below we use 66 seeds
$\ket{\Lambda_\sigma(X)}$, all $\SU(2)\times\Spin(9)$ singlets. Starting from
$\ket S$, the unique $\SU(2)\times\Spin(9)$ singlet in
$\mathcal F$~\cite{Wosiek05,HoppeLundholmTrzetrzelewski} and the seed of
degree zero, we add seeds degree by degree until the
$\ket{P_j}$ form a basis of all singlets of degree
at most five. As one may show from representation theory, this takes 44 seeds.
The remaining 22, fourteen of degree six and eight of degree seven, are the
candidates along which the objective of the optimization described below
decreases fastest, starting from the optimum of a basis whose $\ket{P_j}$ span
all singlets of degree at most four. We keep the $Y^p\ket{\Lambda_\sigma}$ of
degree at most 19 in~$X$, which mix fermionic directions within $\mathcal F_*$.
The companion repository gives the precise definitions.

The ground state extends along the commuting valley and is localized transverse to it, which the
isotropic Gaussian $e^{-as/2}$ captures poorly. We therefore average over
the gauge group an anisotropic Gaussian of width $b^{-1/2}$ along the
valley and $a^{-1/2}$ transverse to it:
\begin{equation}
 \begin{gathered}
  f_{a,b}(s,u,v)=\frac{e^{-as/2}}{4\pi}\int_{S^2}
   e^{(a-b)n^{\mathsf T}Gn/2}\,d\Omega(n),\\
  G=XX^{\mathsf T},\qquad a,b>0.
 \end{gathered}
 \label{eq:invariant-profiles}
\end{equation}
Multiplying the $\ket{P_j}$ by profiles $f_j=f_{a_j,b_j}$ gives the smooth
trial wavefunctions
\begin{equation}
 \ket{\psi_j(X)}=f_j(s,u,v)\ket{P_j(X)}.
 \label{eq:trial-states}
\end{equation}
The full variational basis is given explicitly in the
companion repository.

For fixed $\ket{P_j}$ and profile widths, we expand
$\ket\Psi=\sum_jc_j\ket{\psi_j}$ and minimize the expectation value of
the auxiliary Hamiltonian,
\begin{equation}
 \begin{gathered}
  H_{\mathrm{aux}}\equiv\frac1{16}\sum_{\alpha=1}^{16}Q_\alpha R^2Q_\alpha,\\
  \langle H_{\mathrm{aux}}\rangle_\Psi
   =\frac1{16}\sum_\alpha\frac{\norm{RQ_\alpha\Psi}^2}{\norm{\Psi}^2}.
 \end{gathered}
 \label{eq:auxiliary-hamiltonian}
\end{equation}
The bound of Appendix~\ref{app:fidelity} gives
\begin{equation}
 \begin{gathered}
  F\geq[1-\kappa\langle H_{\mathrm{aux}}\rangle_\Psi]_+,\qquad
  \kappa=\frac{36+8\sqrt{14}}{25},\\
  F\equiv\frac{\me{\Psi}{P_{\ker H}}{\Psi}}{\norm{\Psi}^2},\qquad
  [w]_+=\max\{0,w\},
 \end{gathered}
 \label{eq:variational-fidelity}
\end{equation}
where $P_{\ker H}$ projects onto the zero-energy ground subspace.
Thus minimizing $\langle H_{\mathrm{aux}}\rangle_\Psi$ maximizes the
fidelity lower bound. Defining
\begin{equation}
 \begin{aligned}
  A_{jh}&=\me{\psi_j}{H_{\mathrm{aux}}}{\psi_h}
    =\frac1{16}\sum_\alpha\braket{RQ_\alpha\psi_j}{RQ_\alpha\psi_h},\\
  M_{jh}&=\braket{\psi_j}{\psi_h}
 \end{aligned}
 \label{eq:variational-matrices}
\end{equation}
reduces the optimization to the generalized eigenvalue problem
$Ac=\varepsilon Mc$. Its lowest eigenvalue on the trial span is the
minimum of $\langle H_{\mathrm{aux}}\rangle_\Psi$.
Appendix~\ref{app:ansatz-wick} gives exact expressions for these matrices,
the energy and radial moments using Wick contractions and elementary
angular integrals.

\section{Results}\label{sec:results}

\subsection{The Coulomb-branch EFT}\label{sec:asymptotics}

The Coulomb-branch EFT of the $\SU(2)$ BFSS model describes the slow relative motion of two D0-branes, whose separation is large enough that the degrees of freedom of the W bosons corresponding to stretched strings between them can be treated perturbatively.

More precisely, one performs the change of variables \hypertarget{def:coulomb}{}$X_{a i} = n_a r_i + y_{a i}$ with $n \in S^2, \vec{r} \in \mathbb{R}^9$, $n_a y_{a i} = 0$ and $r_i y_{a i} = 0$. The fermions split alike, $\theta^a_\alpha = n_a \lambda_\alpha + \Theta^a_\alpha$ with $n_a \Theta^a_\alpha = 0$. In terms of the singular value decomposition of $X_{ai}$ with singular values $x_1 \geq x_2 \geq x_3 \geq 0$, one has $r \equiv |\vec r| = x_1$ and $\rho^2 \equiv y^2 = x_2^2 + x_3^2$ and the potential is $u = x_1^2 x_2^2 + x_1^2 x_3^2 + x_2^2 x_3^2$, hence $x_2 = x_3 = 0$ ($y_{a i} = 0$) are vacuum configurations. For $r \gg 1$, the potential becomes $u \simeq \frac12\omega^2 y^2$ (with ``$\simeq$'' denoting the leading order in $1/r$) and $\omega \equiv \sqrt{2} r$, so the bosonic degrees of freedom $y_{a i}$ oscillate with the high frequency $\omega$, and can be integrated out perturbatively (with their partners $\Theta^a_\alpha$) in an expansion around the above vacuum.

At one-loop order in perturbation theory, the ground-state wavefunction $\ket{\Omega}$ factorizes as~\cite{GrafHoppe,FroehlichEtAl,LinYin}
\begin{align}
 \ket{\Omega(X)}&\simeq\frac{\Phi_0(y)}{\sqrt{2\pi}\,r}\,\ket{\psi_{\rm eff}(\vec r)}\otimes\ket{\Omega_F},
 \label{eq:BO}\\
 \ket{\psi_{\rm eff}(\vec r)}&\simeq B\,r^{-9}\ket{\hat r\hat r}.
 \label{eq:psi-eff}
\end{align}
$\Phi_0(y) \simeq (\omega/\pi)^4e^{-\omega y^2/2}$ and $\ket{\Omega_F}$ are the normalized ground states of the heavy $y_{a i}$ and $\Theta^a_\alpha$ respectively, $\int d^{16}y\,\Phi_0^2=1$ and $\braket{\Omega_F}{\Omega_F}=1$. With $\braket{\Omega}{\Omega}=1$, the prefactor makes $\|\psi_{\rm eff}(\vec r)\|^2\,d^9\vec r$ the probability of finding the D0-branes at the relative position in $d^9\vec r$ at large $r$, given $d^{27}X\simeq\frac12r^2\,d\Omega(n)\,d^9\vec r\,d^{16}y$ with $d\Omega(n)$ the area element of $n\in S^2$.

\hypertarget{def:sij}{}Here $\ket{\hat r\hat r}\equiv\hat r_i\hat r_j\ket{s_{ij}}$~\cite{LinYin} (with $\hat{r} \equiv \vec{r}/r$) is a state of the 16 light fermion zero modes $\lambda_\alpha$ in the $\mathbf{44}$ of $\Spin(9)$ normalized so that $\braket{s_{ij}}{s_{kl}}=\tfrac12(\delta_{ik}\delta_{jl}+\delta_{il}\delta_{jk})-\tfrac19\delta_{ij}\delta_{kl}$.

The constant $B$ sets the magnitude of the tail of the unit-normalized ground state; it depends on the core, where the expansion fails, and so cannot be determined from the Coulomb-branch EFT itself. We determine it from the three-point amplitude of~\cite{HerderscheeMaldacena} below.

Given the above leading asymptotic wavefunction, we extract the asymptotics of the radial probability density $p(r) \equiv \langle \delta(r - x_1) \rangle$ and the joint density $p(r, \rho) \equiv \langle \delta(r- x_1) \delta(\rho - \abs{y}) \rangle$. Since $\|\psi_{\rm eff}(\vec r)\|^2=\frac89B^2r^{-18}$ (by $\braket{\hat r\hat r}{\hat r\hat r}=\frac89$) is the probability density of $\vec r$, and at fixed $\vec r$ the $y_{ai}$ are distributed as $|\Phi_0|^2$, we obtain
\begin{align}
 p(r)&\simeq C\,r^{-10},\qquad C=\tfrac89\,\Omega_8B^2,
 \label{eq:tail-benchmark}\\
 p(r,\rho)&\simeq p(r)\,\frac{2\,\omega^8}{7!}\,\rho^{15}e^{-\omega\rho^2},
 \label{eq:joint-density}
\end{align}
with $\Omega_8 = 32 \pi^4/105$ the area of $S^8$. At fixed $r \gg 1$ this gives, for example, $\langle \rho^2 \rangle_r \simeq 8/\omega$ and $\langle \rho^4 \rangle_r \simeq 72/\omega^2$, so the valley (configurations with $p(r, \rho)$ maximal) narrows as $r$ increases.

\subsection{The asymptotic normalization from the three-point amplitude}\label{sec:amplitude}
The coefficient $B$ of~\eqref{eq:psi-eff} controlling the asymptotic normalization of the true ground-state wavefunction can be derived from the three-point amplitude of the $N=2$ bound state decaying into two D0-branes.

As usual, the on-shell three-point amplitude can be defined as the residue of a $2 \to 2$ amplitude through factorization. In our case, factorizing on the ground-state pole at $k^2 = 0$, the appropriate residue defining the on-shell three-point amplitude between the ground state and the two D0-branes is then, with the energy- and momentum-conserving delta functions stripped off, the matrix element $\me{k,\varepsilon}{H_{\rm int}}{\psi_{\rm eff}} \big|_{k^2 = 0}$ times its conjugate. \hypertarget{def:keps}{}Here $\ket{k,\varepsilon}=e^{\ii k\cdot\vec r}\,\varepsilon_i\varepsilon_j\ket{s_{ij}}$ (with the polarization $\varepsilon$ satisfying $\varepsilon \cdot \varepsilon = 0, \varepsilon^* \cdot \varepsilon = 1$) and $H_{\text{int}}$ is the effective interaction Hamiltonian for the light degrees of freedom. We show in Appendix~\ref{app:vertex} that the results of~\cite{HerderscheeMaldacena} (converting to our conventions) for the three-point amplitude give
\begin{equation}
 \me{k,\varepsilon}{H_{\rm int}}{\psi_{\rm eff}}\big|_{k^2=0}=g\,(k\cdot\varepsilon^*)^2,\qquad g=2^{3/4}\pi^2 .
 \label{eq:vertex}
\end{equation}

This matrix element also shows up in the projection of the effective Schr\"odinger equation $(H_0 + H_{\text{int}}) \ket{\psi_{\text{eff}}} = 0$ with $H_0 = -\frac{1}{2}\Delta_{\vec{r}}$ onto $\bra{k,\varepsilon}$,
\begin{equation}
 \braket{k,\varepsilon}{\psi_{\rm eff}}=-\frac{2\,\me{k,\varepsilon}{H_{\rm int}}{\psi_{\rm eff}}}{k^2},
 \label{eq:pole}
\end{equation}
but now at off-shell kinematics. Since a non-polynomial term in $\braket{k,\varepsilon}{\psi_{\rm eff}}$ homogeneous of degree $n$ in $k$ gives a tail $\propto r^{-9-n}$ in $\ket{\psi_{\rm eff}(\vec r)}$, while polynomials give only contact terms, the asymptotic tail of the wavefunction is captured by the pole at $k^2=0$, and, since the source $H_{\rm int}\ket{\psi_{\rm eff}}$ decays sufficiently fast, any dependence of the above matrix element proportional to $k^2$ (hence vanishing on-shell) affects only subleading tails and we may use (\ref{eq:vertex}).

Inserting the on-shell matrix element~\eqref{eq:vertex} into~\eqref{eq:pole} and Fourier transforming, with $\int\frac{d^9k}{(2\pi)^9}\frac{e^{\ii k\cdot\vec r}}{k^2}=\frac{15}{32\pi^4}\,r^{-7}$ and $k_i\to-\ii\partial_i$, gives the tail
\begin{equation}
 \varepsilon^*_i\varepsilon^*_j\braket{s_{ij}}{\psi_{\rm eff}(\vec r)}\simeq\frac{15g}{16\pi^4}\,(\varepsilon^*\!\cdot\partial)^2r^{-7}=\frac{945g}{16\pi^4}\,r^{-9}(\varepsilon^*\!\cdot\hat r)^2 .
\end{equation}
Since $\varepsilon^*_i\varepsilon^*_j\braket{s_{ij}}{\hat r\hat r}=(\varepsilon^*\!\cdot\hat r)^2$, setting this equal to~\eqref{eq:psi-eff} gives
\begin{equation}
  B=945\cdot2^{3/4}/(16\pi^2)
  \label{Beq}
\end{equation}
and, by~\eqref{eq:tail-benchmark}, $C = 1890 \sqrt{2}$.

\subsection{Comparison with the variational state}\label{sec:comparison}

\begin{figure*}[p]
\centering
\includegraphics[width=\figA\textwidth]{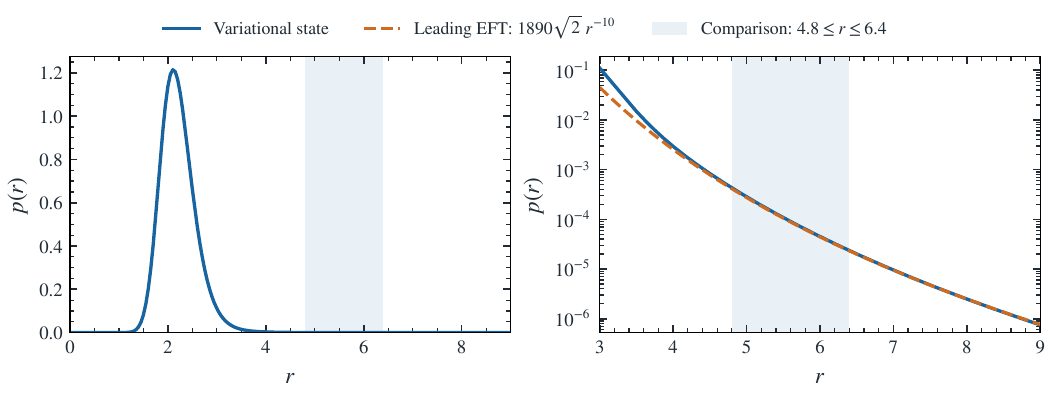}
\caption{Radial density $p(r)$~\eqref{eq:tail-benchmark} of the variational state on a linear (left) and a
logarithmic (right) scale, with the prediction $p\simeq Cr^{-10}$, $C=1890\sqrt2$ (dashed).
The shaded band highlights $\ReportedSlopeRadiusMin{}\leq r\leq\ReportedSlopeRadiusMax{}$.}
\label{fig:eft-radial}
\end{figure*}

\begin{figure*}[p]
\centering
\includegraphics[width=\figB\textwidth]{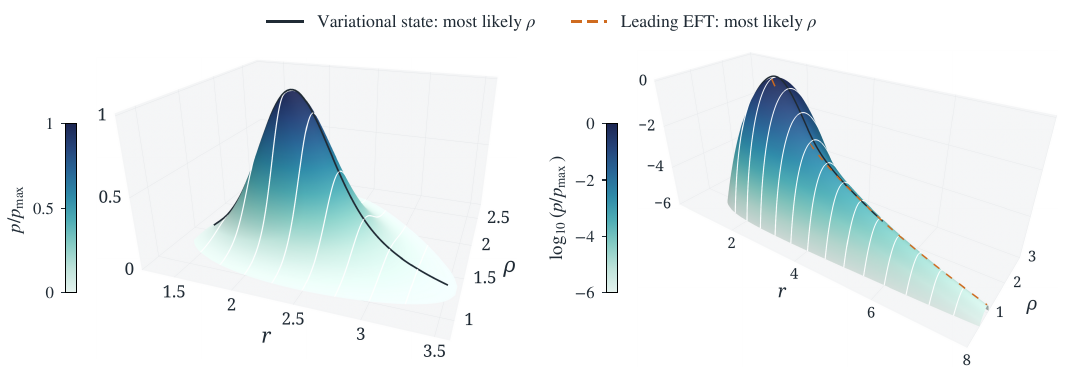}
\caption{Joint density $p(r,\rho)$~\eqref{eq:joint-density} of the variational state relative to its maximum, on a
linear (left) and a logarithmic (right) scale. Black: most likely $\rho$ at each $r$, with its leading
EFT prediction $\rho_*=\sqrt{15/(2\sqrt2\,r)}$ (dashed, right). White: sections at fixed $r$.}
\label{fig:eft-joint}
\end{figure*}

\begin{figure*}[p]
\centering
\includegraphics[width=\figC\textwidth]{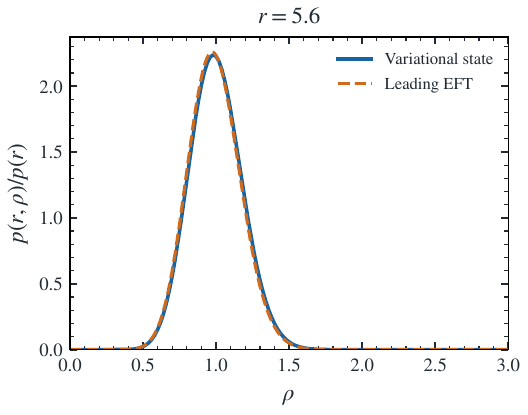}\hspace{.04\textwidth}
\includegraphics[width=\figC\textwidth]{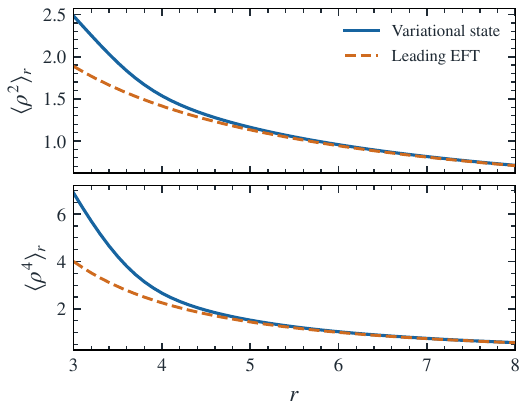}
\caption{Left: conditional density $p(r,\rho)/p(r)$, see~\eqref{eq:tail-benchmark} and~\eqref{eq:joint-density}, of the variational state at
$r=\ReportedTransverseRadius{}$, with the prediction (dashed). Right: its moments $\langle\rho^2\rangle_r$ and
$\langle\rho^4\rangle_r$, from the joint density of Fig.~\ref{fig:eft-joint}, with the leading EFT values $8/\omega$ and $72/\omega^2$ (dashed), $\omega=\sqrt2\,r$.}
\label{fig:eft-transverse}
\end{figure*}

Given the constant $B$ of (\ref{Beq}) derived above, we plot the asymptotic true ground-state predictions for $p(r)$ of~\eqref{eq:tail-benchmark} in Fig.~\ref{fig:eft-radial}, $p(r, \rho)$ of~\eqref{eq:joint-density} in Fig.~\ref{fig:eft-joint} and $p(r, \rho)/p(r)$ at $r = 5.6$ with $\langle\rho^2\rangle_r, \langle\rho^4\rangle_r$ in Fig.~\ref{fig:eft-transverse}, together with those of our variational wavefunction. Here $\langle\cdot\rangle_r$ is the expectation value at fixed $r$, i.e.\ in the conditional density $p(r,\rho)/p(r)$ of~\eqref{eq:joint-density}.

\begin{figure*}[t]
\centering
\includegraphics[width=\textwidth]{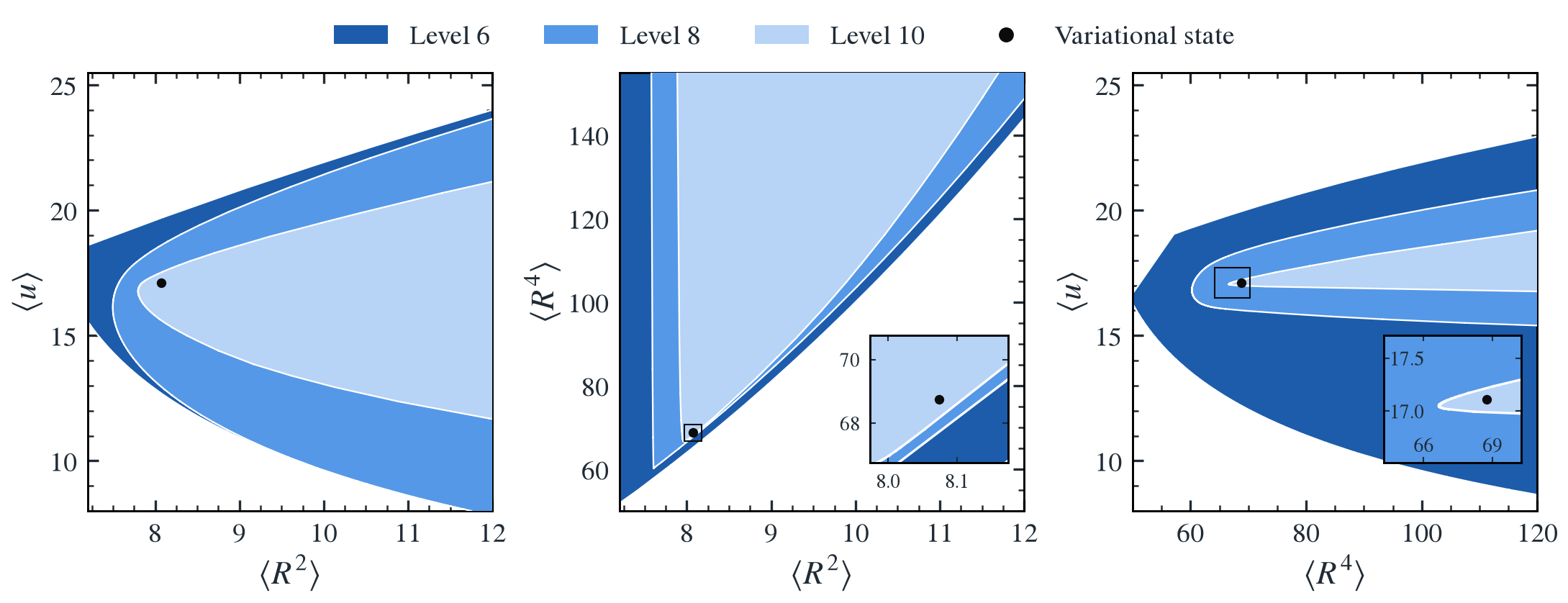}
\caption{Bootstrap regions allowed at levels $6$, $8$ and $10$ (darkest to lightest) in the
planes of $\langle R^2\rangle$, $\langle R^4\rangle$ and $\langle u\rangle$, with $R^2=s$ and $u$ of~\eqref{eq:invariant-coordinates}. Dot: the variational state~\eqref{eq:observables}.
Insets zoom on the region around it.}
\label{fig:bootstrap}
\end{figure*}

At $r$ large enough for the Coulomb-branch EFT to be reliable, the variational state agrees with it
in Figs.~\ref{fig:eft-radial}--\ref{fig:eft-transverse}, both in the radial density and in $\langle\rho^2\rangle_r$, $\langle\rho^4\rangle_r$. Since any truncated
variational state, being a finite sum of Gaussian profiles, eventually decays faster than $p(r)$, the
range of this agreement increases with the size of the basis.

\subsection{The variational ground state, its energy and ground-subspace overlap}\label{sec:wavefunction}
Our variational technique yields a state $\ket{\psi}$, whose explicit coefficients in the truncated variational basis of Subsection~\ref{sec:ansatz} can be found in the companion code~\cite{CompanionCode}. We remind the reader that we work in units where $g_{\mathrm{YM}}=1$. In these units, we have evaluated (in \ReportedMeasurementBits{}-bit ball arithmetic)
\begin{equation}
\begin{aligned}
 E&\leq\ReportedEnergy{}, &
 \langle H_{\mathrm{aux}}\rangle&=\ReportedWeightedResidual{},\\
 \langle T\rangle&=\ReportedKineticEnergy{}, &
 \langle u\rangle&=\ReportedBosonicPotential{},\\
 \langle Y\rangle&=\ReportedYukawa{}, &&\\
 \langle R^2\rangle&=\ReportedRadiusSquared{}, &
 \langle R^4\rangle&=\ReportedRadiusFourth{}.
\end{aligned}
\label{eq:observables}
\end{equation}
Here $E=\langle H\rangle=\|Q_\alpha\psi\|^2$ for each
supercharge~\eqref{eq:supercharge},
$\langle H_{\mathrm{aux}}\rangle=\tfrac1{16}\sum_\alpha\|RQ_\alpha\psi\|^2$
with $R^2=\operatorname{tr}XX^{\mathsf T}$;
$T=-\frac12\Delta_X$,
$u$ and $Y$ are the three terms of $H$ in~\eqref{eq:model-H} (whose values satisfy the virial theorem $2\langle T \rangle = 4 \langle u \rangle + \langle Y \rangle$ to $10^{-6}$).
Parenthetical errors are the rounding to the digits shown; the upper bound on $E$ is rounded up and the lower bound on $F$ down, so both remain valid.

The expectation value of the auxiliary Hamiltonian bounds the ground-subspace overlap $F \equiv \|P_{\ker H}\psi\|^2$ via
$F \geq [1 - \kappa\langle H_{\mathrm{aux}}\rangle]_+$, with $\kappa = (36 + 8 \sqrt{14})/25$ and the bracket denoting the
positive part (Subsection~\ref{sec:fidelity}), so that $F\geq\ReportedFidelityLowerBoundValue{}$.
The Rayleigh--Ritz method converges quickly in our trial space: a basis of only
\ConvergenceTermCount{} functions generated from \ConvergenceSeedCount{} seeds already gives
$F\geq\ConvergenceFidelity{}$, and the whole computation runs on a laptop.
The overlap rigorously controls the deviation of any bounded observable (continuous in the Hilbert-space norm) from its true ground-state value. The operators whose expectation values we evaluated above are unbounded, so we do not have rigorous control over their expectation values in the above sense. We can still estimate the deviations of their expectation values from those in the true ground state (for example, one can multiply $R^2$ by a compact indicator to make it bounded and estimate the remainder).

The above results are consistent with our comparatively loose bounds from a
matrix-quantum-mechanics bootstrap~\cite{Lin20,HanHartnollKruthoff,KazakovZheng,LinD0Bootstrap,LinTASI},
shown in Fig.~\ref{fig:bootstrap}, certifying $\langle R^2 \rangle \geq 7.798$ and $\langle R^4 \rangle \geq 66.67$, below our $\langle R^2 \rangle = \ReportedRadiusSquared{}$ and $\langle R^4 \rangle = \ReportedRadiusFourth{}$. Our bootstrap constrains the moments of a ground state $\ket{\Omega}$ annihilated by $H$, the supercharges and the generators of $\SU(2) \times \Spin(9)$, that is, $\langle \Omega | \mathcal{O} N | \Omega \rangle = 0$ for each operator $\mathcal{O}$ and $N$ the generators. As usual, the bounds then follow from further imposing positivity of the matrix elements $\langle \Omega| \mathcal{O}^\dagger \mathcal{O} | \Omega\rangle \geq 0$. In Fig.~\ref{fig:bootstrap}, we consider operators $\mathcal{O}$ up to level 10, where each $X_{ai}$ has level 2, each momentum $P_{ai}=-\ii\partial_{X_{ai}}$ level 4 and each $\theta^a_\alpha$ level 3. Note that we omit operators whose $\mathcal{O}^\dagger \mathcal{O}$ has no finite expectation value, such as $\mathcal{O}_{ai} = X_{ai} R^4$. The code that produces these bounds is in the companion repository.

\section{Summary and discussion}
\label{sec:discussion}
We used the Rayleigh--Ritz method to study the ground state of $\SU(2)$ BFSS, leveraging an enormous reduction in the degrees of freedom coming from representation theory. The method had to be nontrivially adapted to $\SU(2)$ BFSS as the theory is not gapped. We overcame this by introducing an auxiliary operator with a gapped spectrum that has the same ground states as the physical Hamiltonian. The gap of this operator was related to a gap of a spinor operator via SUSY, and this gap was bounded rigorously. This allowed us to verify that our variational state overlapped the true ground state in excess of \ReportedFidelity{}. In this state of high fidelity and energy $E\,g_{\mathrm{YM}}^{-2/3}\leq\ReportedEnergy{}$, we accurately computed various moments, consistent with our comparatively loose bootstrap bounds. Moreover, we matched predictions on the asymptotic behavior of the ground-state wavefunction that were extracted from the three-point amplitude of \cite{HerderscheeMaldacena}, namely that the radial probability density decays like $p(r) \sim 1890\sqrt{2} \, r^{-10}$ for large $r$.

Several extensions suggest themselves. The most immediate one is to study time-independent observables at higher $N$, for which a first target could be to observe the large-$N$ scaling of moments such as $\langle R^2 \rangle \gtrsim N^{\frac{5}{3}}, \, N \gg 1$~\cite{Polchinski99}. Since the reduction from representation theory that worked for $\SU(2)$ does not extend to higher $N$, nonlinear variational approaches (e.g.\ neural ans\"atze~\cite{CarleoTroyer,PfauFermiNet,HermannPauliNet,HanHartnoll,RinaldiMatrixModels}) may be necessary.

Furthermore, it is an outstanding challenge to study time-dependent observables, especially at large $N$, where scattering on the Coulomb branch gets related to M-theory scattering via the BFSS conjecture~\cite{BFSS,Polchinski99,MaldacenaStrings24,LinTASI}. Perhaps $N=2$ is sufficiently large and one might observe some of the key features expected of such scattering.

Lastly, we mention that the $\SU(N)$ BFSS model is expected to possess a large number of resonances buried in the continuum, which become long-lived and correspond to microstates of black holes at large $N$. It would be quite interesting to see these resonances even in the $\SU(2)$ case, carrying out a study analogous to that of \cite{BalthazarRodriguezYin}, who, through an adapted L\"uscher-type method, discovered many metastable resonances in toy models possessing many of the key features.

\paragraph*{Note added.} We have been informed by Kabir Bajaj and Henry Lin that they have independently determined moments of the $\SU(2)$ BFSS ground state using stronger bootstrap bounds, and have also accurately determined the asymptotic tail of the wavefunction. We thank them for sharing their results with us.

\section*{Acknowledgments}
We thank Kabir Bajaj, Henry Lin and Zechuan Zheng for discussions on the bootstrap bounds.
We thank Sam Christian for useful discussions.
JS and XY thank the RIKEN Interdisciplinary Theoretical and Mathematical
Sciences Program (iTHEMS) and the Yukawa Institute for Theoretical Physics
(YITP) at Kyoto University for hospitality and a stimulating atmosphere, in
particular during the iTHEMS-YITP workshop `Resurgence, Matrix model, String field theory and
Instantons' (YITP-W-26-11). This work is supported by DOE grant
DE-SC0007870. GPT 5.6 Sol, GPT 6.0 Astra and Claude Opus 5.5 were
used to write code and to assist with writing and editing the manuscript.

The computations in this paper were run on the FASRC Cannon cluster supported by the FAS Division of Science Research Computing Group at Harvard University.

\bibliography{references}

\begin{thebibliography}{42}%
\makeatletter
\providecommand \@ifxundefined [1]{%
 \@ifx{#1\undefined}
}%
\providecommand \@ifnum [1]{%
 \ifnum #1\expandafter \@firstoftwo
 \else \expandafter \@secondoftwo
 \fi
}%
\providecommand \@ifx [1]{%
 \ifx #1\expandafter \@firstoftwo
 \else \expandafter \@secondoftwo
 \fi
}%
\providecommand \natexlab [1]{#1}%
\providecommand \enquote  [1]{``#1''}%
\providecommand \bibnamefont  [1]{#1}%
\providecommand \bibfnamefont [1]{#1}%
\providecommand \citenamefont [1]{#1}%
\providecommand \href@noop [0]{\@secondoftwo}%
\providecommand \href [0]{\begingroup \@sanitize@url \@href}%
\providecommand \@href[1]{\@@startlink{#1}\@@href}%
\providecommand \@@href[1]{\endgroup#1\@@endlink}%
\providecommand \@sanitize@url [0]{\catcode `\\12\catcode `\$12\catcode
  `\&12\catcode `\#12\catcode `\^12\catcode `\_12\catcode `\%12\relax}%
\providecommand \@@startlink[1]{}%
\providecommand \@@endlink[0]{}%
\providecommand \url  [0]{\begingroup\@sanitize@url \@url }%
\providecommand \@url [1]{\endgroup\@href {#1}{\urlprefix }}%
\providecommand \urlprefix  [0]{URL }%
\providecommand \Eprint [0]{\href }%
\providecommand \doibase [0]{https://doi.org/}%
\providecommand \selectlanguage [0]{\@gobble}%
\providecommand \bibinfo  [0]{\@secondoftwo}%
\providecommand \bibfield  [0]{\@secondoftwo}%
\providecommand \translation [1]{[#1]}%
\providecommand \BibitemOpen [0]{}%
\providecommand \bibitemStop [0]{}%
\providecommand \bibitemNoStop [0]{.\EOS\space}%
\providecommand \EOS [0]{\spacefactor3000\relax}%
\providecommand \BibitemShut  [1]{\csname bibitem#1\endcsname}%
\let\auto@bib@innerbib\@empty
\bibitem [{\citenamefont {Banks}\ \emph {et~al.}(1997)\citenamefont {Banks},
  \citenamefont {Fischler}, \citenamefont {Shenker},\ and\ \citenamefont
  {Susskind}}]{BFSS}%
  \BibitemOpen
  \bibfield  {author} {\bibinfo {author} {\bibfnamefont {T.}~\bibnamefont
  {Banks}}, \bibinfo {author} {\bibfnamefont {W.}~\bibnamefont {Fischler}},
  \bibinfo {author} {\bibfnamefont {S.~H.}\ \bibnamefont {Shenker}},\ and\
  \bibinfo {author} {\bibfnamefont {L.}~\bibnamefont {Susskind}},\ }\bibfield
  {title} {\bibinfo {title} {{M theory as a matrix model: A conjecture}},\
  }\href@noop {} {\bibfield  {journal} {\bibinfo  {journal} {Phys. Rev. D}\
  }\textbf {\bibinfo {volume} {55}},\ \bibinfo {pages} {5112} (\bibinfo {year}
  {1997})},\ \Eprint {https://arxiv.org/abs/hep-th/9610043}
  {arXiv:hep-th/9610043} \BibitemShut {NoStop}%
\bibitem [{\citenamefont {Polchinski}(1999)}]{Polchinski99}%
  \BibitemOpen
  \bibfield  {author} {\bibinfo {author} {\bibfnamefont {J.}~\bibnamefont
  {Polchinski}},\ }\bibfield  {title} {\bibinfo {title} {{M-theory and the
  light cone}},\ }\href@noop {} {\bibfield  {journal} {\bibinfo  {journal}
  {Prog. Theor. Phys. Suppl.}\ }\textbf {\bibinfo {volume} {134}},\ \bibinfo
  {pages} {158} (\bibinfo {year} {1999})},\ \Eprint
  {https://arxiv.org/abs/hep-th/9903165} {arXiv:hep-th/9903165} \BibitemShut
  {NoStop}%
\bibitem [{\citenamefont {Maldacena}(2024)}]{MaldacenaStrings24}%
  \BibitemOpen
  \bibfield  {author} {\bibinfo {author} {\bibfnamefont {J.}~\bibnamefont
  {Maldacena}},\ }\href@noop {} {\bibinfo {title} {{The BFSS conjecture, a
  review}}},\ \bibinfo {howpublished} {talk at Strings 2024, CERN, Geneva,
  \href{https://member.ipmu.jp/yuji.tachikawa/stringsmirrors/2024/Maldacena.pdf}{slides}}
  (\bibinfo {year} {2024})\BibitemShut {NoStop}%
\bibitem [{\citenamefont {Lin}(2025)}]{LinTASI}%
  \BibitemOpen
  \bibfield  {author} {\bibinfo {author} {\bibfnamefont {H.~W.}\ \bibnamefont
  {Lin}},\ }\href@noop {} {\bibinfo {title} {{TASI lectures on Matrix Theory
  from a modern viewpoint}}} (\bibinfo {year} {2025}),\ \Eprint
  {https://arxiv.org/abs/2508.20970} {arXiv:2508.20970} \BibitemShut {NoStop}%
\bibitem [{\citenamefont {Maldacena}(1998)}]{Maldacena97}%
  \BibitemOpen
  \bibfield  {author} {\bibinfo {author} {\bibfnamefont {J.~M.}\ \bibnamefont
  {Maldacena}},\ }\bibfield  {title} {\bibinfo {title} {{The large $N$ limit of
  superconformal field theories and supergravity}},\ }\href@noop {} {\bibfield
  {journal} {\bibinfo  {journal} {Adv. Theor. Math. Phys.}\ }\textbf {\bibinfo
  {volume} {2}},\ \bibinfo {pages} {231} (\bibinfo {year} {1998})},\ \Eprint
  {https://arxiv.org/abs/hep-th/9711200} {arXiv:hep-th/9711200} \BibitemShut
  {NoStop}%
\bibitem [{\citenamefont {Balasubramanian}\ \emph {et~al.}(1998)\citenamefont
  {Balasubramanian}, \citenamefont {Gopakumar},\ and\ \citenamefont
  {Larsen}}]{BGL}%
  \BibitemOpen
  \bibfield  {author} {\bibinfo {author} {\bibfnamefont {V.}~\bibnamefont
  {Balasubramanian}}, \bibinfo {author} {\bibfnamefont {R.}~\bibnamefont
  {Gopakumar}},\ and\ \bibinfo {author} {\bibfnamefont {F.}~\bibnamefont
  {Larsen}},\ }\bibfield  {title} {\bibinfo {title} {{Gauge theory, geometry
  and the large $N$ limit}},\ }\href@noop {} {\bibfield  {journal} {\bibinfo
  {journal} {Nucl. Phys. B}\ }\textbf {\bibinfo {volume} {526}},\ \bibinfo
  {pages} {415} (\bibinfo {year} {1998})},\ \Eprint
  {https://arxiv.org/abs/hep-th/9712077} {arXiv:hep-th/9712077} \BibitemShut
  {NoStop}%
\bibitem [{\citenamefont {Itzhaki}\ \emph {et~al.}(1998)\citenamefont
  {Itzhaki}, \citenamefont {Maldacena}, \citenamefont {Sonnenschein},\ and\
  \citenamefont {Yankielowicz}}]{IMSY}%
  \BibitemOpen
  \bibfield  {author} {\bibinfo {author} {\bibfnamefont {N.}~\bibnamefont
  {Itzhaki}}, \bibinfo {author} {\bibfnamefont {J.~M.}\ \bibnamefont
  {Maldacena}}, \bibinfo {author} {\bibfnamefont {J.}~\bibnamefont
  {Sonnenschein}},\ and\ \bibinfo {author} {\bibfnamefont {S.}~\bibnamefont
  {Yankielowicz}},\ }\bibfield  {title} {\bibinfo {title} {{Supergravity and
  the large $N$ limit of theories with sixteen supercharges}},\ }\href@noop {}
  {\bibfield  {journal} {\bibinfo  {journal} {Phys. Rev. D}\ }\textbf {\bibinfo
  {volume} {58}},\ \bibinfo {pages} {046004} (\bibinfo {year} {1998})},\
  \Eprint {https://arxiv.org/abs/hep-th/9802042} {arXiv:hep-th/9802042}
  \BibitemShut {NoStop}%
\bibitem [{\citenamefont {Sethi}\ and\ \citenamefont
  {Stern}(1998)}]{SethiStern}%
  \BibitemOpen
  \bibfield  {author} {\bibinfo {author} {\bibfnamefont {S.}~\bibnamefont
  {Sethi}}\ and\ \bibinfo {author} {\bibfnamefont {M.}~\bibnamefont {Stern}},\
  }\bibfield  {title} {\bibinfo {title} {{D-brane bound states redux}},\
  }\href@noop {} {\bibfield  {journal} {\bibinfo  {journal} {Commun. Math.
  Phys.}\ }\textbf {\bibinfo {volume} {194}},\ \bibinfo {pages} {675} (\bibinfo
  {year} {1998})},\ \Eprint {https://arxiv.org/abs/hep-th/9705046}
  {arXiv:hep-th/9705046} \BibitemShut {NoStop}%
\bibitem [{\citenamefont {Yi}(1997)}]{Yi97}%
  \BibitemOpen
  \bibfield  {author} {\bibinfo {author} {\bibfnamefont {P.}~\bibnamefont
  {Yi}},\ }\bibfield  {title} {\bibinfo {title} {{Witten index and threshold
  bound states of D-branes}},\ }\href@noop {} {\bibfield  {journal} {\bibinfo
  {journal} {Nucl. Phys. B}\ }\textbf {\bibinfo {volume} {505}},\ \bibinfo
  {pages} {307} (\bibinfo {year} {1997})},\ \Eprint
  {https://arxiv.org/abs/hep-th/9704098} {arXiv:hep-th/9704098} \BibitemShut
  {NoStop}%
\bibitem [{\citenamefont {Sethi}\ and\ \citenamefont
  {Stern}(2000)}]{SethiStern00}%
  \BibitemOpen
  \bibfield  {author} {\bibinfo {author} {\bibfnamefont {S.}~\bibnamefont
  {Sethi}}\ and\ \bibinfo {author} {\bibfnamefont {M.}~\bibnamefont {Stern}},\
  }\bibfield  {title} {\bibinfo {title} {{Invariance theorems for
  supersymmetric Yang--Mills theories}},\ }\href@noop {} {\bibfield  {journal}
  {\bibinfo  {journal} {Adv. Theor. Math. Phys.}\ }\textbf {\bibinfo {volume}
  {4}},\ \bibinfo {pages} {487} (\bibinfo {year} {2000})},\ \Eprint
  {https://arxiv.org/abs/hep-th/0001189} {arXiv:hep-th/0001189} \BibitemShut
  {NoStop}%
\bibitem [{\citenamefont {Hasler}\ and\ \citenamefont
  {Hoppe}(2002{\natexlab{a}})}]{HaslerHoppe}%
  \BibitemOpen
  \bibfield  {author} {\bibinfo {author} {\bibfnamefont {D.}~\bibnamefont
  {Hasler}}\ and\ \bibinfo {author} {\bibfnamefont {J.}~\bibnamefont {Hoppe}},\
  }\href@noop {} {\bibinfo {title} {{Zero energy states of reduced super
  Yang--Mills theories in $d+1=4$, $6$ and $10$ dimensions are necessarily
  $\mathrm{Spin}(d)$ invariant}}} (\bibinfo {year} {2002}{\natexlab{a}}),\
  \Eprint {https://arxiv.org/abs/hep-th/0211226} {arXiv:hep-th/0211226}
  \BibitemShut {NoStop}%
\bibitem [{\citenamefont {Porrati}\ and\ \citenamefont
  {Rozenberg}(1998)}]{PorratiRozenberg}%
  \BibitemOpen
  \bibfield  {author} {\bibinfo {author} {\bibfnamefont {M.}~\bibnamefont
  {Porrati}}\ and\ \bibinfo {author} {\bibfnamefont {A.}~\bibnamefont
  {Rozenberg}},\ }\bibfield  {title} {\bibinfo {title} {{Bound states at
  threshold in supersymmetric quantum mechanics}},\ }\href@noop {} {\bibfield
  {journal} {\bibinfo  {journal} {Nucl. Phys. B}\ }\textbf {\bibinfo {volume}
  {515}},\ \bibinfo {pages} {184} (\bibinfo {year} {1998})},\ \Eprint
  {https://arxiv.org/abs/hep-th/9708119} {arXiv:hep-th/9708119} \BibitemShut
  {NoStop}%
\bibitem [{\citenamefont {Green}\ and\ \citenamefont
  {Gutperle}(1998)}]{GreenGutperle}%
  \BibitemOpen
  \bibfield  {author} {\bibinfo {author} {\bibfnamefont {M.~B.}\ \bibnamefont
  {Green}}\ and\ \bibinfo {author} {\bibfnamefont {M.}~\bibnamefont
  {Gutperle}},\ }\bibfield  {title} {\bibinfo {title} {{D-particle bound states
  and the D-instanton measure}},\ }\href@noop {} {\bibfield  {journal}
  {\bibinfo  {journal} {JHEP}\ }\textbf {\bibinfo {volume} {1998}}\bibfield
  {number} {\bibinfo  {number} { (01)},\ \bibinfo {pages} {005}},\ }\Eprint
  {https://arxiv.org/abs/hep-th/9711107} {arXiv:hep-th/9711107} \BibitemShut
  {NoStop}%
\bibitem [{\citenamefont {Moore}\ \emph {et~al.}(2000)\citenamefont {Moore},
  \citenamefont {Nekrasov},\ and\ \citenamefont {Shatashvili}}]{MNS}%
  \BibitemOpen
  \bibfield  {author} {\bibinfo {author} {\bibfnamefont {G.~W.}\ \bibnamefont
  {Moore}}, \bibinfo {author} {\bibfnamefont {N.}~\bibnamefont {Nekrasov}},\
  and\ \bibinfo {author} {\bibfnamefont {S.}~\bibnamefont {Shatashvili}},\
  }\bibfield  {title} {\bibinfo {title} {{D-particle bound states and
  generalized instantons}},\ }\href@noop {} {\bibfield  {journal} {\bibinfo
  {journal} {Commun. Math. Phys.}\ }\textbf {\bibinfo {volume} {209}},\
  \bibinfo {pages} {77} (\bibinfo {year} {2000})},\ \Eprint
  {https://arxiv.org/abs/hep-th/9803265} {arXiv:hep-th/9803265} \BibitemShut
  {NoStop}%
\bibitem [{\citenamefont {Kac}\ and\ \citenamefont {Smilga}(2000)}]{KacSmilga}%
  \BibitemOpen
  \bibfield  {author} {\bibinfo {author} {\bibfnamefont {V.~G.}\ \bibnamefont
  {Kac}}\ and\ \bibinfo {author} {\bibfnamefont {A.~V.}\ \bibnamefont
  {Smilga}},\ }\bibfield  {title} {\bibinfo {title} {{Normalized vacuum states
  in $\mathcal N=4$ supersymmetric Yang--Mills quantum mechanics with any gauge
  group}},\ }\href@noop {} {\bibfield  {journal} {\bibinfo  {journal} {Nucl.
  Phys. B}\ }\textbf {\bibinfo {volume} {571}},\ \bibinfo {pages} {515}
  (\bibinfo {year} {2000})},\ \Eprint {https://arxiv.org/abs/hep-th/9908096}
  {arXiv:hep-th/9908096} \BibitemShut {NoStop}%
\bibitem [{\citenamefont {Staudacher}(2000)}]{Staudacher}%
  \BibitemOpen
  \bibfield  {author} {\bibinfo {author} {\bibfnamefont {M.}~\bibnamefont
  {Staudacher}},\ }\bibfield  {title} {\bibinfo {title} {{Bulk Witten indices
  and the number of normalizable ground states in supersymmetric quantum
  mechanics of orthogonal, symplectic and exceptional groups}},\ }\href@noop {}
  {\bibfield  {journal} {\bibinfo  {journal} {Phys. Lett. B}\ }\textbf
  {\bibinfo {volume} {488}},\ \bibinfo {pages} {194} (\bibinfo {year}
  {2000})},\ \Eprint {https://arxiv.org/abs/hep-th/0006234}
  {arXiv:hep-th/0006234} \BibitemShut {NoStop}%
\bibitem [{\citenamefont {Lee}\ and\ \citenamefont {Yi}(2016)}]{LeeYi}%
  \BibitemOpen
  \bibfield  {author} {\bibinfo {author} {\bibfnamefont {S.-J.}\ \bibnamefont
  {Lee}}\ and\ \bibinfo {author} {\bibfnamefont {P.}~\bibnamefont {Yi}},\
  }\bibfield  {title} {\bibinfo {title} {{Witten index for noncompact
  dynamics}},\ }\href@noop {} {\bibfield  {journal} {\bibinfo  {journal}
  {JHEP}\ }\textbf {\bibinfo {volume} {2016}}\bibfield  {number} {\bibinfo
  {number} { (06)},\ \bibinfo {pages} {089}},\ }\Eprint
  {https://arxiv.org/abs/1602.03530} {arXiv:1602.03530} \BibitemShut {NoStop}%
\bibitem [{\citenamefont {{OpenAI}}()}]{OpenAIBFSS}%
  \BibitemOpen
  \bibfield  {author} {\bibinfo {author} {\bibnamefont {{OpenAI}}},\
  }\href@noop {} {\bibinfo {title} {{The unique threshold bound state of the
  SU($N$) BFSS model}}},\ \bibinfo {howpublished} {preprint (September 24,
  2026),
  \url{https://github.com/openai/math/blob/main/preprints/The-unique-threshold-bound-state-of-the-SU-N-BFSS-model-September-24-2026/paper.pdf}}\BibitemShut
  {NoStop}%
\bibitem [{\citenamefont {Halpern}\ and\ \citenamefont
  {Schwartz}(1998)}]{HalpernSchwartz}%
  \BibitemOpen
  \bibfield  {author} {\bibinfo {author} {\bibfnamefont {M.~B.}\ \bibnamefont
  {Halpern}}\ and\ \bibinfo {author} {\bibfnamefont {C.}~\bibnamefont
  {Schwartz}},\ }\bibfield  {title} {\bibinfo {title} {{Asymptotic search for
  ground states of SU(2) matrix theory}},\ }\href@noop {} {\bibfield  {journal}
  {\bibinfo  {journal} {Int. J. Mod. Phys. A}\ }\textbf {\bibinfo {volume}
  {13}},\ \bibinfo {pages} {4367} (\bibinfo {year} {1998})},\ \Eprint
  {https://arxiv.org/abs/hep-th/9712133} {arXiv:hep-th/9712133} \BibitemShut
  {NoStop}%
\bibitem [{\citenamefont {Konechny}(1998)}]{Konechny}%
  \BibitemOpen
  \bibfield  {author} {\bibinfo {author} {\bibfnamefont {A.}~\bibnamefont
  {Konechny}},\ }\bibfield  {title} {\bibinfo {title} {{On asymptotic
  Hamiltonian for SU($N$) matrix theory}},\ }\href@noop {} {\bibfield
  {journal} {\bibinfo  {journal} {JHEP}\ }\textbf {\bibinfo {volume}
  {1998}}\bibfield  {number} {\bibinfo  {number} { (10)},\ \bibinfo {pages}
  {018}},\ }\Eprint {https://arxiv.org/abs/hep-th/9805046}
  {arXiv:hep-th/9805046} \BibitemShut {NoStop}%
\bibitem [{\citenamefont {Graf}\ and\ \citenamefont {Hoppe}(1998)}]{GrafHoppe}%
  \BibitemOpen
  \bibfield  {author} {\bibinfo {author} {\bibfnamefont {G.~M.}\ \bibnamefont
  {Graf}}\ and\ \bibinfo {author} {\bibfnamefont {J.}~\bibnamefont {Hoppe}},\
  }\href@noop {} {\bibinfo {title} {{Asymptotic ground state for
  ten-dimensional reduced supersymmetric SU(2) Yang--Mills theory}}} (\bibinfo
  {year} {1998}),\ \Eprint {https://arxiv.org/abs/hep-th/9805080}
  {arXiv:hep-th/9805080} \BibitemShut {NoStop}%
\bibitem [{\citenamefont {Fr{\"o}hlich}\ \emph {et~al.}(2000)\citenamefont
  {Fr{\"o}hlich}, \citenamefont {Graf}, \citenamefont {Hasler}, \citenamefont
  {Hoppe},\ and\ \citenamefont {Yau}}]{FroehlichEtAl}%
  \BibitemOpen
  \bibfield  {author} {\bibinfo {author} {\bibfnamefont {J.}~\bibnamefont
  {Fr{\"o}hlich}}, \bibinfo {author} {\bibfnamefont {G.~M.}\ \bibnamefont
  {Graf}}, \bibinfo {author} {\bibfnamefont {D.}~\bibnamefont {Hasler}},
  \bibinfo {author} {\bibfnamefont {J.}~\bibnamefont {Hoppe}},\ and\ \bibinfo
  {author} {\bibfnamefont {S.-T.}\ \bibnamefont {Yau}},\ }\bibfield  {title}
  {\bibinfo {title} {{Asymptotic form of zero energy wave functions in
  supersymmetric matrix models}},\ }\href@noop {} {\bibfield  {journal}
  {\bibinfo  {journal} {Nucl. Phys. B}\ }\textbf {\bibinfo {volume} {567}},\
  \bibinfo {pages} {231} (\bibinfo {year} {2000})},\ \Eprint
  {https://arxiv.org/abs/hep-th/9904182} {arXiv:hep-th/9904182} \BibitemShut
  {NoStop}%
\bibitem [{\citenamefont {Hasler}\ and\ \citenamefont
  {Hoppe}(2002{\natexlab{b}})}]{HaslerHoppeAsym}%
  \BibitemOpen
  \bibfield  {author} {\bibinfo {author} {\bibfnamefont {D.}~\bibnamefont
  {Hasler}}\ and\ \bibinfo {author} {\bibfnamefont {J.}~\bibnamefont {Hoppe}},\
  }\href@noop {} {\bibinfo {title} {{Asymptotic factorization of the ground
  state for SU($N$) invariant supersymmetric matrix models}}} (\bibinfo {year}
  {2002}{\natexlab{b}}),\ \Eprint {https://arxiv.org/abs/hep-th/0206043}
  {arXiv:hep-th/0206043} \BibitemShut {NoStop}%
\bibitem [{\citenamefont {Hoppe}\ \emph {et~al.}(2009)\citenamefont {Hoppe},
  \citenamefont {Lundholm},\ and\ \citenamefont
  {Trzetrzelewski}}]{HoppeLundholmTrzetrzelewski}%
  \BibitemOpen
  \bibfield  {author} {\bibinfo {author} {\bibfnamefont {J.}~\bibnamefont
  {Hoppe}}, \bibinfo {author} {\bibfnamefont {D.}~\bibnamefont {Lundholm}},\
  and\ \bibinfo {author} {\bibfnamefont {M.}~\bibnamefont {Trzetrzelewski}},\
  }\bibfield  {title} {\bibinfo {title} {{Construction of the zero-energy state
  of SU(2)-matrix theory: near the origin}},\ }\href@noop {} {\bibfield
  {journal} {\bibinfo  {journal} {Nucl. Phys. B}\ }\textbf {\bibinfo {volume}
  {817}},\ \bibinfo {pages} {155} (\bibinfo {year} {2009})},\ \Eprint
  {https://arxiv.org/abs/0809.5270} {arXiv:0809.5270} \BibitemShut {NoStop}%
\bibitem [{\citenamefont {Michishita}\ and\ \citenamefont
  {Trzetrzelewski}(2013)}]{MichishitaTrzetrzelewski}%
  \BibitemOpen
  \bibfield  {author} {\bibinfo {author} {\bibfnamefont {Y.}~\bibnamefont
  {Michishita}}\ and\ \bibinfo {author} {\bibfnamefont {M.}~\bibnamefont
  {Trzetrzelewski}},\ }\bibfield  {title} {\bibinfo {title} {{Towards the
  ground state of the supermembrane}},\ }\href@noop {} {\bibfield  {journal}
  {\bibinfo  {journal} {Nucl. Phys. B}\ }\textbf {\bibinfo {volume} {868}},\
  \bibinfo {pages} {539} (\bibinfo {year} {2013})},\ \Eprint
  {https://arxiv.org/abs/1208.4978} {arXiv:1208.4978} \BibitemShut {NoStop}%
\bibitem [{\citenamefont {Lin}\ and\ \citenamefont {Yin}(2015)}]{LinYin}%
  \BibitemOpen
  \bibfield  {author} {\bibinfo {author} {\bibfnamefont {Y.-H.}\ \bibnamefont
  {Lin}}\ and\ \bibinfo {author} {\bibfnamefont {X.}~\bibnamefont {Yin}},\
  }\bibfield  {title} {\bibinfo {title} {{On the ground state wave function of
  matrix theory}},\ }\href@noop {} {\bibfield  {journal} {\bibinfo  {journal}
  {JHEP}\ }\textbf {\bibinfo {volume} {2015}}\bibfield  {number} {\bibinfo
  {number} { (11)},\ \bibinfo {pages} {027}},\ }\Eprint
  {https://arxiv.org/abs/1402.0055} {arXiv:1402.0055} \BibitemShut {NoStop}%
\bibitem [{\citenamefont {Lin}(2020)}]{Lin20}%
  \BibitemOpen
  \bibfield  {author} {\bibinfo {author} {\bibfnamefont {H.~W.}\ \bibnamefont
  {Lin}},\ }\bibfield  {title} {\bibinfo {title} {{Bootstraps to strings:
  solving random matrix models with positivity}},\ }\href@noop {} {\bibfield
  {journal} {\bibinfo  {journal} {JHEP}\ }\textbf {\bibinfo {volume}
  {2020}}\bibfield  {number} {\bibinfo  {number} { (06)},\ \bibinfo {pages}
  {090}},\ }\Eprint {https://arxiv.org/abs/2002.08387} {arXiv:2002.08387}
  \BibitemShut {NoStop}%
\bibitem [{\citenamefont {Han}\ \emph {et~al.}(2020)\citenamefont {Han},
  \citenamefont {Hartnoll},\ and\ \citenamefont
  {Kruthoff}}]{HanHartnollKruthoff}%
  \BibitemOpen
  \bibfield  {author} {\bibinfo {author} {\bibfnamefont {X.}~\bibnamefont
  {Han}}, \bibinfo {author} {\bibfnamefont {S.~A.}\ \bibnamefont {Hartnoll}},\
  and\ \bibinfo {author} {\bibfnamefont {J.}~\bibnamefont {Kruthoff}},\
  }\bibfield  {title} {\bibinfo {title} {{Bootstrapping matrix quantum
  mechanics}},\ }\href@noop {} {\bibfield  {journal} {\bibinfo  {journal}
  {Phys. Rev. Lett.}\ }\textbf {\bibinfo {volume} {125}},\ \bibinfo {pages}
  {041601} (\bibinfo {year} {2020})},\ \Eprint
  {https://arxiv.org/abs/2004.10212} {arXiv:2004.10212} \BibitemShut {NoStop}%
\bibitem [{\citenamefont {Kazakov}\ and\ \citenamefont
  {Zheng}(2022)}]{KazakovZheng}%
  \BibitemOpen
  \bibfield  {author} {\bibinfo {author} {\bibfnamefont {V.}~\bibnamefont
  {Kazakov}}\ and\ \bibinfo {author} {\bibfnamefont {Z.}~\bibnamefont
  {Zheng}},\ }\bibfield  {title} {\bibinfo {title} {{Analytic and numerical
  bootstrap for one-matrix model and ``unsolvable'' two-matrix model}},\
  }\href@noop {} {\bibfield  {journal} {\bibinfo  {journal} {JHEP}\ }\textbf
  {\bibinfo {volume} {2022}}\bibfield  {number} {\bibinfo  {number} { (06)},\
  \bibinfo {pages} {030}},\ }\Eprint {https://arxiv.org/abs/2108.04830}
  {arXiv:2108.04830} \BibitemShut {NoStop}%
\bibitem [{\citenamefont {Lin}(2023)}]{LinD0Bootstrap}%
  \BibitemOpen
  \bibfield  {author} {\bibinfo {author} {\bibfnamefont {H.~W.}\ \bibnamefont
  {Lin}},\ }\bibfield  {title} {\bibinfo {title} {{Bootstrap bounds on D0-brane
  quantum mechanics}},\ }\href@noop {} {\bibfield  {journal} {\bibinfo
  {journal} {JHEP}\ }\textbf {\bibinfo {volume} {2023}}\bibfield  {number}
  {\bibinfo  {number} { (06)},\ \bibinfo {pages} {038}},\ }\Eprint
  {https://arxiv.org/abs/2302.04416} {arXiv:2302.04416} \BibitemShut {NoStop}%
\bibitem [{\citenamefont {Herderschee}\ and\ \citenamefont
  {Maldacena}(2023)}]{HerderscheeMaldacena}%
  \BibitemOpen
  \bibfield  {author} {\bibinfo {author} {\bibfnamefont {A.}~\bibnamefont
  {Herderschee}}\ and\ \bibinfo {author} {\bibfnamefont {J.}~\bibnamefont
  {Maldacena}},\ }\href@noop {} {\bibinfo {title} {{Three point amplitudes in
  matrix theory}}} (\bibinfo {year} {2023}),\ \Eprint
  {https://arxiv.org/abs/2312.12592} {arXiv:2312.12592} \BibitemShut {NoStop}%
\bibitem [{Com()}]{CompanionCode}%
  \BibitemOpen
  \href@noop {} {\bibinfo {title} {{Companion code}}},\ \bibinfo {note}
  {\ReportedCodeLink{}; the reported state is \ReportedStateLink{}; the
  bootstrap code is \ReportedBootstrapLink{}}\BibitemShut {NoStop}%
\bibitem [{\citenamefont {Fulton}\ and\ \citenamefont
  {Harris}(1991)}]{FultonHarris}%
  \BibitemOpen
  \bibfield  {author} {\bibinfo {author} {\bibfnamefont {W.}~\bibnamefont
  {Fulton}}\ and\ \bibinfo {author} {\bibfnamefont {J.}~\bibnamefont
  {Harris}},\ }\href@noop {} {\emph {\bibinfo {title} {{Representation Theory:
  A First Course}}}},\ \bibinfo {series} {Graduate Texts in Mathematics}, Vol.\
  \bibinfo {volume} {129}\ (\bibinfo  {publisher} {Springer},\ \bibinfo
  {address} {New York},\ \bibinfo {year} {1991})\BibitemShut {NoStop}%
\bibitem [{\citenamefont {Wosiek}(2005)}]{Wosiek05}%
  \BibitemOpen
  \bibfield  {author} {\bibinfo {author} {\bibfnamefont {J.}~\bibnamefont
  {Wosiek}},\ }\bibfield  {title} {\bibinfo {title} {{On the SO(9) structure of
  supersymmetric Yang--Mills quantum mechanics}},\ }\href@noop {} {\bibfield
  {journal} {\bibinfo  {journal} {Phys. Lett. B}\ }\textbf {\bibinfo {volume}
  {619}},\ \bibinfo {pages} {171} (\bibinfo {year} {2005})},\ \Eprint
  {https://arxiv.org/abs/hep-th/0503236} {arXiv:hep-th/0503236} \BibitemShut
  {NoStop}%
\bibitem [{\citenamefont {Carleo}\ and\ \citenamefont
  {Troyer}(2017)}]{CarleoTroyer}%
  \BibitemOpen
  \bibfield  {author} {\bibinfo {author} {\bibfnamefont {G.}~\bibnamefont
  {Carleo}}\ and\ \bibinfo {author} {\bibfnamefont {M.}~\bibnamefont
  {Troyer}},\ }\bibfield  {title} {\bibinfo {title} {{Solving the quantum
  many-body problem with artificial neural networks}},\ }\href@noop {}
  {\bibfield  {journal} {\bibinfo  {journal} {Science}\ }\textbf {\bibinfo
  {volume} {355}},\ \bibinfo {pages} {602} (\bibinfo {year} {2017})},\ \Eprint
  {https://arxiv.org/abs/1606.02318} {arXiv:1606.02318} \BibitemShut {NoStop}%
\bibitem [{\citenamefont {Pfau}\ \emph {et~al.}(2020)\citenamefont {Pfau},
  \citenamefont {Spencer}, \citenamefont {Matthews},\ and\ \citenamefont
  {Foulkes}}]{PfauFermiNet}%
  \BibitemOpen
  \bibfield  {author} {\bibinfo {author} {\bibfnamefont {D.}~\bibnamefont
  {Pfau}}, \bibinfo {author} {\bibfnamefont {J.~S.}\ \bibnamefont {Spencer}},
  \bibinfo {author} {\bibfnamefont {A.~G. D.~G.}\ \bibnamefont {Matthews}},\
  and\ \bibinfo {author} {\bibfnamefont {W.~M.~C.}\ \bibnamefont {Foulkes}},\
  }\bibfield  {title} {\bibinfo {title} {{Ab initio solution of the
  many-electron Schr{\"o}dinger equation with deep neural networks}},\
  }\href@noop {} {\bibfield  {journal} {\bibinfo  {journal} {Phys. Rev.
  Research}\ }\textbf {\bibinfo {volume} {2}},\ \bibinfo {pages} {033429}
  (\bibinfo {year} {2020})},\ \Eprint {https://arxiv.org/abs/1909.02487}
  {arXiv:1909.02487} \BibitemShut {NoStop}%
\bibitem [{\citenamefont {Hermann}\ \emph {et~al.}(2020)\citenamefont
  {Hermann}, \citenamefont {Sch{\"a}tzle},\ and\ \citenamefont
  {No{\'e}}}]{HermannPauliNet}%
  \BibitemOpen
  \bibfield  {author} {\bibinfo {author} {\bibfnamefont {J.}~\bibnamefont
  {Hermann}}, \bibinfo {author} {\bibfnamefont {Z.}~\bibnamefont
  {Sch{\"a}tzle}},\ and\ \bibinfo {author} {\bibfnamefont {F.}~\bibnamefont
  {No{\'e}}},\ }\bibfield  {title} {\bibinfo {title} {{Deep-neural-network
  solution of the electronic Schr{\"o}dinger equation}},\ }\href@noop {}
  {\bibfield  {journal} {\bibinfo  {journal} {Nature Chem.}\ }\textbf {\bibinfo
  {volume} {12}},\ \bibinfo {pages} {891} (\bibinfo {year} {2020})},\ \Eprint
  {https://arxiv.org/abs/1909.08423} {arXiv:1909.08423} \BibitemShut {NoStop}%
\bibitem [{\citenamefont {Han}\ and\ \citenamefont
  {Hartnoll}(2020)}]{HanHartnoll}%
  \BibitemOpen
  \bibfield  {author} {\bibinfo {author} {\bibfnamefont {X.}~\bibnamefont
  {Han}}\ and\ \bibinfo {author} {\bibfnamefont {S.~A.}\ \bibnamefont
  {Hartnoll}},\ }\bibfield  {title} {\bibinfo {title} {{Deep quantum geometry
  of matrices}},\ }\href@noop {} {\bibfield  {journal} {\bibinfo  {journal}
  {Phys. Rev. X}\ }\textbf {\bibinfo {volume} {10}},\ \bibinfo {pages} {011069}
  (\bibinfo {year} {2020})},\ \Eprint {https://arxiv.org/abs/1906.08781}
  {arXiv:1906.08781} \BibitemShut {NoStop}%
\bibitem [{\citenamefont {Rinaldi}\ \emph {et~al.}(2022)\citenamefont
  {Rinaldi}, \citenamefont {Han}, \citenamefont {Hassan}, \citenamefont {Feng},
  \citenamefont {Nori}, \citenamefont {McGuigan},\ and\ \citenamefont
  {Hanada}}]{RinaldiMatrixModels}%
  \BibitemOpen
  \bibfield  {author} {\bibinfo {author} {\bibfnamefont {E.}~\bibnamefont
  {Rinaldi}}, \bibinfo {author} {\bibfnamefont {X.}~\bibnamefont {Han}},
  \bibinfo {author} {\bibfnamefont {M.}~\bibnamefont {Hassan}}, \bibinfo
  {author} {\bibfnamefont {Y.}~\bibnamefont {Feng}}, \bibinfo {author}
  {\bibfnamefont {F.}~\bibnamefont {Nori}}, \bibinfo {author} {\bibfnamefont
  {M.}~\bibnamefont {McGuigan}},\ and\ \bibinfo {author} {\bibfnamefont
  {M.}~\bibnamefont {Hanada}},\ }\bibfield  {title} {\bibinfo {title}
  {{Matrix-model simulations using quantum computing, deep learning, and
  lattice Monte Carlo}},\ }\href@noop {} {\bibfield  {journal} {\bibinfo
  {journal} {PRX Quantum}\ }\textbf {\bibinfo {volume} {3}},\ \bibinfo {pages}
  {010324} (\bibinfo {year} {2022})},\ \Eprint
  {https://arxiv.org/abs/2108.02942} {arXiv:2108.02942} \BibitemShut {NoStop}%
\bibitem [{\citenamefont {Balthazar}\ \emph {et~al.}(2019)\citenamefont
  {Balthazar}, \citenamefont {Rodriguez},\ and\ \citenamefont
  {Yin}}]{BalthazarRodriguezYin}%
  \BibitemOpen
  \bibfield  {author} {\bibinfo {author} {\bibfnamefont {B.}~\bibnamefont
  {Balthazar}}, \bibinfo {author} {\bibfnamefont {V.~A.}\ \bibnamefont
  {Rodriguez}},\ and\ \bibinfo {author} {\bibfnamefont {X.}~\bibnamefont
  {Yin}},\ }\bibfield  {title} {\bibinfo {title} {{Hamiltonian truncation study
  of supersymmetric quantum mechanics: S-matrix and metastable states}},\
  }\href@noop {} {\bibfield  {journal} {\bibinfo  {journal} {JHEP}\ }\textbf
  {\bibinfo {volume} {2019}}\bibfield  {number} {\bibinfo  {number} { (08)},\
  \bibinfo {pages} {100}},\ }\Eprint {https://arxiv.org/abs/1610.07275}
  {arXiv:1610.07275} \BibitemShut {NoStop}%
\bibitem [{\citenamefont {Yin}()}]{YinNotes}%
  \BibitemOpen
  \bibfield  {author} {\bibinfo {author} {\bibfnamefont {X.}~\bibnamefont
  {Yin}},\ }\href@noop {} {\bibinfo {title} {{Foundations of String Theory}}},\
  \bibinfo {howpublished}
  {\url{https://github.com/xiyin137/stringbook}}\BibitemShut {NoStop}%
\bibitem [{\citenamefont {Edelman}\ and\ \citenamefont
  {Rao}(2005)}]{EdelmanRao}%
  \BibitemOpen
  \bibfield  {author} {\bibinfo {author} {\bibfnamefont {A.}~\bibnamefont
  {Edelman}}\ and\ \bibinfo {author} {\bibfnamefont {N.~R.}\ \bibnamefont
  {Rao}},\ }\bibfield  {title} {\bibinfo {title} {{Random matrix theory}},\
  }\href {https://doi.org/10.1017/S0962492904000236} {\bibfield  {journal}
  {\bibinfo  {journal} {Acta Numer.}\ }\textbf {\bibinfo {volume} {14}},\
  \bibinfo {pages} {233} (\bibinfo {year} {2005})}\BibitemShut {NoStop}%
\end{thebibliography}%

\clearpage
\onecolumngrid
\makeatletter
\global\let\set@footnotewidth\set@footnotewidth@one
\global\expandafter\let\csname combine@insert@1\endcsname\combine@insert@@ne
\global\count\footins\@m
\makeatother
\appendix
\numberwithin{equation}{section}
\numberwithin{theorem}{section}
\numberwithin{proposition}{section}
\renewcommand{\theHequation}{\thesection.\arabic{equation}}
\renewcommand{\theHtheorem}{\thesection.\arabic{theorem}}
\section{Fermion and gamma-matrix conventions}
\label{sec:fermions}

Here we give the explicit basis of $\mathcal F$ and the real gamma matrices
$\Gamma_i$ of~\eqref{eq:model-H} in which the coefficients of our
variational state (Subsection~\ref{sec:wavefunction}) are given.

\subsection{Multi-indices}
\label{sec:multi-indices}

A multi-index is a tuple $I=(i_1,\ldots,i_d)$ with
components $i_j=1,\ldots,n_j$. Ordering the tuples lexicographically,
with the last component varying fastest, we identify each tuple with its
position in this order, a flat index $i_{\rm flat}\in\mathbb Z$:
\begin{equation}
 (i_1,\ldots,i_d)\sim i_{\rm flat}
 \iff
 i_{\rm flat}=1+\sum_{j=1}^d(i_j-1)\,n_{j+1}\cdots n_d .
 \label{eq:multi-index}
\end{equation}
This pairs the tuples one-to-one with $i_{\rm flat}=1,\ldots,n_1\cdots n_d$,
with inverse
$i_j=1+\bigl(\lfloor(i_{\rm flat}-1)/(n_{j+1}\cdots n_d)\rfloor\bmod n_j\bigr)$.
We then freely write $I$ in either form, $I=(i_1,\ldots,i_d)$
or $I=i_{\rm flat}$, without ambiguity,
and in subscripts also juxtapose the components of the
tuple. Nested multi-indices
flatten consistently, $((i_1,i_2),i_3)=(i_1,i_2,i_3)$, and tensor products
of matrices are Kronecker products,
$(M\otimes M')_{(i,i'),(j,j')}=M_{ij}M'_{i'j'}$. The companion code uses
the same order, numbering every index from 0.

\subsection{Majorana operators and gamma matrices}
\label{sec:gammas}

The real gamma matrices are
\begin{equation}
 \begin{gathered}
 \Gamma_i=\operatorname{Re}H_i\otimes I_2-\operatorname{Im}H_i\otimes\varepsilon,\qquad
 \Gamma_{3+m}=\operatorname{Re}B_m\otimes\sigma_3+\operatorname{Im}B_m\otimes\sigma_1,\\
 H_i=I_4\otimes h_i,\qquad B_m=\rho_m\otimes\varepsilon,\qquad
 \varepsilon=\ii\sigma_2,
 \end{gathered}
 \label{eq:real-gammas}
\end{equation}
where $h=(\sigma_1,\sigma_2,-\sigma_3)$ in terms of the
Pauli matrices and, with $E_{AB}=e_{AB}-e_{BA}$,
\begin{equation}
 \begin{aligned}
 \rho_1&=E_{12}+E_{34},\qquad &\rho_2&=\ii(E_{12}-E_{34}),\\
 \rho_3&=E_{13}-E_{24},\qquad &\rho_4&=\ii(E_{13}+E_{24}),\\
 \rho_5&=E_{14}+E_{23},\qquad &\rho_6&=\ii(E_{14}-E_{23}).
 \end{aligned}
\end{equation}
Hermiticity of $H_i$ and symmetry of $B_m$ prove real symmetry of these
matrices. The elementary two- and four-dimensional products give
$H_iH_j+H_jH_i=2\delta_{ij}$,
$H_iB_m+B_m\overline H_i=0$, and
$B_m\overline B_n+B_n\overline B_m=2\delta_{mn}$; these imply the
Clifford relations for the real matrices $\Gamma_i$.
The generators~\eqref{eq:gauge-generators}
and~\eqref{eq:fermion-rotations} act on the Majorana operators by
\begin{equation}
 [\mathcal G_{ab},\theta^c_\alpha]=\delta_{bc}\theta^a_\alpha-\delta_{ac}\theta^b_\alpha,\qquad
 [\mathcal R_{ij},\theta^a_\alpha]=-\tfrac12\sum_\beta(\Gamma_i\Gamma_j)_{\alpha\beta}\theta^a_\beta.
 \label{eq:majorana-symmetry-action}
\end{equation}

\subsection{Complex fermions}
\label{sec:complex-fermions}

We choose complex fermions
adapted to the stabilizer $K_4\times\SO(6)$~\eqref{eq:diagonal-stabilizer} of
Section~\ref{sec:symmetry-reduction}: under the rotations
$\SO(3)\times\SO(6)\subset\SO(9)$ of the first three and the last six
directions, the spinor index splits into a spin index of
$\Spin(3)\simeq\SU(2)$ and a flavor index of $\Spin(6)\simeq\SU(4)$, so
that the $\SO(6)$ of the stabilizer acts on the flavor alone and $K_4$ only
on color and spin.

Writing the real spinor index as the multi-index
$\alpha=(r,s)$, with $r=1,\ldots,8$ and $s=1,2$, we pair the 48 Majorana
operators of Section~\ref{sec:model} in $s$ into the 24 complex fermions
\begin{equation}
 f^a_r=\frac{\theta^a_{r,1}+\ii\theta^a_{r,2}}{\sqrt2},\qquad
 \{f^a_r,f^{b\dagger}_{r'}\}=\delta_{ab}\delta_{rr'},\qquad
 \{f^a_r,f^b_{r'}\}=0,
 \label{eq:majorana-polarization}
\end{equation}
and split their spinor index as $r=(A,\rho)$, with the flavor index
$A=1,\ldots,4$ and the spin index $\rho=1,2$.
By~\eqref{eq:real-gammas} and~\eqref{eq:majorana-symmetry-action}, the
color rotations and $\Spin(3)\times\Spin(6)$ conserve the fermion number
$\sum_{a,r}f^{a\dagger}_rf^a_r$, acting on the $f^a_r$ as
\begin{equation}
 \begin{gathered}
 [\mathcal G_{ab},f^c_r]=\delta_{bc}f^a_r-\delta_{ac}f^b_r,\\
 [\mathcal R_{ij},f^a_{A\rho}]
   =-\tfrac12\sum_{\rho'}(h_ih_j)_{\rho\rho'}f^a_{A\rho'},\qquad i,j\leq3,\\
 [\mathcal R_{3+m,3+n},f^a_{A\rho}]
   =\tfrac12\sum_{A'}(\rho_m\overline\rho_n)_{AA'}f^a_{A'\rho}.
 \end{gathered}
 \label{eq:fermion-symmetry-action}
\end{equation}
Thus $\Spin(3)$ acts on the spin index $\rho$ alone, with $\rho=1,2$ for
down and up, and $\Spin(6)$ on the flavor index $A$ alone, through the
fifteen matrices $\tfrac12\rho_m\overline\rho_n$, $m<n$, which span
$\mathfrak{su}(4)$. The remaining rotations $\mathcal R_{i,3+m}$,
$i\leq3$, change the fermion number by~$\pm2$. The stabilizer fixes the
pairing up to $f\leftrightarrow f^\dagger$: the real $16\times16$ matrices
commuting with $\Spin(9)$ are multiples of $\one$, while those commuting
with $\Spin(3)\times\Spin(6)$ are spanned by $\one$ and the complex
structure $J=I_8\otimes(-\varepsilon)$ of the pairing.

\subsection{Occupation basis}
\label{sec:occupation-basis}

For the occupation basis we use the spherical basis of the adjoint,
\begin{equation}
 f^a_{A\rho}=\sum_{c=1}^3U_{ac}f_{Ac\rho},\qquad
 U=\begin{pmatrix}
 1/\sqrt2&0&-1/\sqrt2\\
 \ii/\sqrt2&0&\ii/\sqrt2\\
 0&1&0
 \end{pmatrix},
 \label{eq:color-polarization}
\end{equation}
in which $f_{Ac\rho}$ with $c=1,2,3$ has weight $-1,0,1$ under color
rotations about the 3-axis; the creation operators transform with the
complex-conjugate coefficients. Explicitly,
by~\eqref{eq:fermion-symmetry-action}, the color and spin rotations about
the 3-axis act as
\begin{equation}
 [\mathcal G_{12},f_{Ac\rho}]=\ii(c-2)f_{Ac\rho},\qquad
 [\mathcal R_{12},f_{Ac\rho}]=\ii(\rho-\tfrac32)f_{Ac\rho}.
\end{equation}
As $U$ is unitary, the modes $f_m$,
$m=(A,c,\rho)=(A,\ell)$ with $\ell=(c,\rho)$, again satisfy
$\{f_m,f^\dagger_{m'}\}=\delta_{mm'}$ and
$\{f_m,f_{m'}\}=0$. For occupation numbers $n_m\in\{0,1\}$, the companion
code labels the occupation states by binary numerals:
\begin{equation}
 |M\rangle=(f^\dagger_1)^{n_1}\cdots(f^\dagger_{24})^{n_{24}}|0\rangle,
 \qquad M=(n_{24}\cdots n_1)_2.
 \label{eq:fock-mask}
\end{equation}
The occupation states $|M\rangle$ are orthonormal. The action of $f_m$
or $f_m^\dagger$ has sign
$(-1)^{n_1+\cdots+n_{m-1}}$ when the
annihilated mode is occupied or the created mode is unoccupied,
respectively, and is zero otherwise. In an operator product, the
rightmost factor acts first.

\clearpage
\section{Proof of the ground-space fidelity bound}
\label{app:fidelity}

Here we prove the fidelity bound~\eqref{eq:variational-fidelity}.
Recall from Section~\ref{sec:model} that on the gauge-invariant space $\HH$
each supercharge~\eqref{eq:supercharge} squares to the Hamiltonian,
$Q_\alpha^2=H$, so that $\me{\eta}{H}{\eta}=\norm{Q_\alpha\eta}^2$, and that
the $Q_\alpha$ transform as the spinor $\mathbf{16}$ of $\Spin(9)$, while $H$
and $R^2=\tr XX^{\mathsf T}$ are invariant.

The bound rests on the Hardy-type inequality~\eqref{app-eq:hardy} for the states of $\HH$
transforming in the $\mathbf{16}$, such as $Q_\alpha\ket{\psi}$ for a singlet $\ket{\psi}$.
Spinor states carry angular momentum, so their energy controls their
inverse-square radius. We prove the inequality after deriving
Theorem~\ref{app-thm:fidelity} from it.

\begin{lemma}\label{app-lem:hardy}
Let $\HH_S$ be the spinor sector of $\HH$, spanned by all copies
of the $\mathbf{16}$. Every $\ket{\eta}\in\HH_S$ satisfies
\begin{equation}\label{app-eq:hardy}
 \me{\eta}{H}{\eta}\geq\kappa^{-1}\norm{R^{-1}\eta}^2,
\end{equation}
with $\kappa=(36+8\sqrt{14})/25$ as in~\eqref{eq:variational-fidelity}.
\end{lemma}

Since $Q_\alpha^2=H$, the weight $1-F$ that a normalized singlet
$\ket{\psi}$ carries outside $\ker H$ is
$\me{Q_\alpha\psi}{H^{-1}}{Q_\alpha\psi}$ for each $\alpha$. With
Cauchy--Schwarz, the lemma bounds it by $\kappa\norm{RQ_\alpha\psi}^2$, whose
average over $\alpha$ is $\kappa\langle H_{\mathrm{aux}}\rangle_\psi$ for the
auxiliary Hamiltonian~\eqref{eq:auxiliary-hamiltonian}.

\begin{theorem}\label{app-thm:fidelity}
With $H_{\mathrm{aux}}\equiv\tfrac1{16}\sum_\alpha Q_\alpha R^2Q_\alpha$, every normalized
$\Spin(9)$ singlet $\ket{\psi}\in\HH$ satisfies
\begin{equation}\label{app-eq:fidelity}
 F=\me{\psi}{P_{\ker H}}{\psi}
 \geq[1-\kappa\langle H_{\mathrm{aux}}\rangle_\psi]_+.
\end{equation}
Equivalently, restricted to singlets, $H_{\mathrm{aux}}$ has a gap of at least
$\kappa^{-1}=(9-2\sqrt{14})/4$ above its zero modes, which are those of $H$.
\end{theorem}

\begin{proof}[Proof of Theorem~\ref{app-thm:fidelity}]
We regulate $H^{-1}$ as $(H+\varepsilon)^{-1}$, $\varepsilon>0$.
Since $\ket{\psi}$ is a singlet, $\ket{\chi}=Q_\alpha\ket{\psi}$ lies in
$\HH_S$, and so does $\ket{\eta}=(H+\varepsilon)^{-1}\ket{\chi}$, as $H$ is
rotation invariant. As
$\braket{\chi}{\eta}=\me{\eta}{H}{\eta}+\varepsilon\norm{\eta}^2$, the lemma
gives
$\braket{\chi}{\eta}=\braket{R\chi}{R^{-1}\eta}
\leq\norm{R\chi}\sqrt{\kappa\braket{\chi}{\eta}}$, so that
\begin{equation}\label{app-eq:resolvent}
 \me{\chi}{(H+\varepsilon)^{-1}}{\chi}
 \leq\kappa\norm{R\chi}^2.
\end{equation}
Let $\rho_\psi(E)$ be the spectral density of $\ket{\psi}$ at energies
$E>0$, so that
$\me{\psi}{f(H)}{\psi}=Ff(0)+\int_{0^+}^\infty dE\,\rho_\psi(E)f(E)$.
Then $Q_\alpha^2=H$ gives
\begin{equation}\label{app-eq:spectral-limit}
 \me{\chi}{(H+\varepsilon)^{-1}}{\chi}
 =\int_{0^+}^\infty dE\,\rho_\psi(E)\frac{E}{E+\varepsilon}
 \xrightarrow{\varepsilon\to0}1-F,
\end{equation}
so~\eqref{app-eq:resolvent} gives
$1-F\leq\kappa\norm{RQ_\alpha\psi}^2$ for every $\alpha$.\footnote{Applied
to $e^{-tH/2}Q_\alpha\ket{\psi}$, $t\geq0$, the same argument gives
$F\geq Z(t)-\kappa\norm{Re^{-tH/2}Q_\alpha\psi}^2$, where
$Z(t)=\me{\psi}{e^{-tH}}{\psi}\geq e^{-t\langle H\rangle_\psi}$ by Jensen's
inequality. This bound saturates to the exact fidelity as $t\to\infty$,
although we do not use it in this work.}
Averaging over $\alpha$ gives $1-F\leq\kappa\langle H_{\mathrm{aux}}\rangle_\psi$.
No spectral gap of $H$ is needed, since $E/(E+\varepsilon)\to1$ for every
$E>0$, however small. Finally,
$\langle H_{\mathrm{aux}}\rangle_\psi=\tfrac1{16}\sum_\alpha\norm{RQ_\alpha\psi}^2$
vanishes exactly
when $\me{\psi}{H}{\psi}=\norm{Q_\alpha\psi}^2$ does, so the zero modes of
$H_{\mathrm{aux}}$ are those of $H$. For singlets orthogonal to them, $F=0$,
and~\eqref{app-eq:fidelity} gives the gap.
\end{proof}

\begin{proof}[Proof of Lemma~\ref{app-lem:hardy}]
Write $\HH_S\cong\mathbf{16}\otimes\mathcal M$, with $\mathcal M$
the multiplicity space. Since $H$ and $R$ commute with rotations, Schur's
lemma implies that they act on $\mathcal M$ only. For
$\ket{\eta}=\sum_b\ket{s_b}\otimes\ket{m_b}$, with $\ket{s_b}$ an orthonormal
basis of the $\mathbf{16}$, both sides of~\eqref{app-eq:hardy} are therefore
sums over $b$ of terms depending on $\ket{m_b}$ alone. It suffices to
prove~\eqref{app-eq:hardy} for $\ket{\eta}=\ket{s_+}\otimes\ket{m}$ with a
fixed unit spinor $\ket{s_+}$ and arbitrary $\ket{m}\in\mathcal M$. We take
$\ket{s_+}$ of highest weight, with eigenvalue $\frac12$ under each of the
angular momenta $J_{12},J_{34},J_{56},J_{78}$, where
$J_{ij}=-\ii(\mathcal R_{ij}+\mathcal S_{ij})$ generates the rotations of
Section~\ref{sec:model}. Then $J=J_{12}+J_{34}+J_{56}+J_{78}$ satisfies
$J\ket{\eta}=2\ket{\eta}$.

For $i<j$, define
$\tau^{ij}_\alpha=\sum_{a,\beta}(\Upsilon^{ij}_a)_{\alpha\beta}\theta^a_\beta$
with the real coefficient matrices
\begin{equation}\label{app-eq:A}
 \Upsilon_a^{ij}=\frac{X_{ai}\Gamma_j-X_{aj}\Gamma_i}{16}
       +\frac1{56}\sum_{k\ne i,j}X_{ak}\Gamma_k\Gamma_i\Gamma_j.
\end{equation}
They satisfy
\begin{equation}\label{app-eq:rotation}
 \sum_\alpha\{Q_\alpha,\tau^{ij}_\alpha\}=J_{ij}.
\end{equation}
Indeed, in the anticommutator with the derivative term
$-\ii\sum_{a,k}C^\alpha_{ak}\partial_{X_{ak}}$ of $Q_\alpha$
in~\eqref{eq:supercharge}, the Majorana anticommutator and the derivative of
$\tau^{ij}_\alpha$ leave $-\ii\mathcal S_{ij}$ and $-\ii\mathcal R_{ij}$,
respectively, by
\begin{equation}
 \tr(\Gamma_k\Upsilon^{ij}_a)=X_{ai}\delta_{jk}-X_{aj}\delta_{ik},\qquad
 \sum_k\Gamma_k\partial_{X_{ak}}\Upsilon^{ij}_a
 =\Bigl(\frac2{16}+\frac7{56}\Bigr)\Gamma_i\Gamma_j
 =\frac14\Gamma_i\Gamma_j.
\end{equation}
The other term $V_\alpha$ of $Q_\alpha$ is linear in the Majorana operators,
with coefficient matrices in the span of the $\Gamma_k\Gamma_l$. Its
contribution is thus a sum of traces of products of three or five gamma
matrices. These vanish: after cancelling repeated factors, such a product has
an odd number of distinct factors, at most five, and reverses sign under
conjugation by a gamma matrix with an unused index.

Let $\tau_\alpha=\tau^{12}_\alpha+\tau^{34}_\alpha
+\tau^{56}_\alpha+\tau^{78}_\alpha$ and
$\Xi_\alpha=\sum_{a,k}X_{ak}C^\alpha_{ak}$. By the Majorana anticommutator and
the orthogonality of products of distinct gamma matrices under the trace, each
plane $(i,j)$ contributes
$\sum_a[(X_{ai}^2+X_{aj}^2)/32+\sum_{k\ne i,j}X_{ak}^2/392]$ to
$\sum_\alpha\tau_\alpha^2$, with no cross terms, since the products
in~\eqref{app-eq:A} differ between planes. Hence
\begin{equation}\label{app-eq:norms}
 \sum_\alpha\tau_\alpha^2=\frac{61R^2-45\sum_aX_{a9}^2}{1568},\qquad
 \sum_\alpha\Xi_\alpha^2=8R^2.
\end{equation}
The traceless part of $\sum_aX_{ai}X_{aj}R^{-4}$ transforms in the
$\mathbf{44}$ of symmetric traceless tensors. This is absent from
$\mathbf{16}\otimes\mathbf{16}=\mathbf1\oplus\mathbf9\oplus\mathbf{36}
\oplus\mathbf{84}\oplus\mathbf{126}$, the antisymmetric tensors of rank at most four, so its
expectation value in $\ket{\eta}$ vanishes. Thus
$\me{\eta}{\sum_aX_{a9}^2R^{-4}}{\eta}=\frac19\norm{R^{-1}\eta}^2$,
and~\eqref{app-eq:norms} gives
\begin{equation}\label{app-eq:C}
 \sum_\alpha\norm{R^{-2}\tau_\alpha\eta}^2
 =\frac{61-45/9}{1568}\norm{R^{-1}\eta}^2=\frac1{28}\norm{R^{-1}\eta}^2,
 \qquad
 \sum_\alpha\norm{R^{-2}\Xi_\alpha\eta}^2=8\norm{R^{-1}\eta}^2.
\end{equation}

Finally, $[Q_\alpha,R^{-2}]=2\ii R^{-4}\Xi_\alpha$
by~\eqref{eq:supercharge}, so summing~\eqref{app-eq:rotation} over the four
planes gives
\begin{equation}\label{app-eq:weighted-rotation}
 \sum_\alpha\{Q_\alpha,R^{-2}\tau_\alpha\}
 =R^{-2}J+2\ii R^{-4}\sum_\alpha\Xi_\alpha\tau_\alpha.
\end{equation}
In $\ket{\eta}$, the term $R^{-2}J$ has expectation value
$2\norm{R^{-1}\eta}^2$. Cauchy--Schwarz with
$\sum_\alpha\norm{Q_\alpha\eta}^2=16\me{\eta}{H}{\eta}$
and~\eqref{app-eq:C} bounds the anticommutator by the energy, while the
commutator term is a fraction of $\norm{R^{-1}\eta}^2$ and can be absorbed:
\begin{equation}\label{app-eq:weighted-bound}
\begin{aligned}
 2\norm{R^{-1}\eta}^2
 &\leq2\Bigl|\sum_\alpha\braket{Q_\alpha\eta}{R^{-2}\tau_\alpha\eta}\Bigr|
       +2\Bigl|\sum_\alpha
          \braket{R^{-2}\Xi_\alpha\eta}{R^{-2}\tau_\alpha\eta}\Bigr|\\
 &\leq\frac4{\sqrt7}\sqrt{\me{\eta}{H}{\eta}}\,\norm{R^{-1}\eta}
       +2\sqrt{\frac27}\,\norm{R^{-1}\eta}^2.
\end{aligned}
\end{equation}
Since $\sqrt{2/7}<1$, this rearranges to~\eqref{app-eq:hardy} with
$\kappa=4/(\sqrt7-\sqrt2)^2=(36+8\sqrt{14})/25$.
\end{proof}

\clearpage
\section{The on-shell matrix element}
\label{app:tail-normalization}\label{app:vertex}

Here we derive~\eqref{eq:vertex} from~\cite{HerderscheeMaldacena}. In the holographic dual of the BFSS model~\cite{IMSY,BGL,Polchinski99,MaldacenaStrings24,LinTASI,YinNotes}, a bound state of $N$ D0-branes carries momentum $N/R_-$ along the M-theory circle $y\sim y+2\pi R_-$, which becomes lightlike at the asymptotic boundary, and the energy and momenta of our Hamiltonian~\eqref{eq:model-H} at $g_{\mathrm{YM}}=1$ are the M-theory energy and transverse momenta in units of $(2\pi)^{-2/3}M_{11}^2R_-$ and $(2\pi)^{-1/3}M_{11}$, respectively, with $M_{11}$ the eleven-dimensional Planck mass. A result in M-theory units is thus converted to ours by setting $(2\pi)^{-1/3}M_{11}=1$ and $(2\pi)^{-2/3}M_{11}^2R_-=1$, i.e.\ $M_{11}=(2\pi)^{1/3}$ and $R_-=1$.

For D0-brane charges $N_1,N_2\to N_3$ and KK-momenta $n_a$ along a further compact direction $x^9$ of radius $R_9$, Eq.~(15) of~\cite{HerderscheeMaldacena} (v4, with $l_p=M_{11}^{-1}$) gives the amplitude between one-particle states with $\braket{p}{p'}=(2\pi)^8\delta^8(\vec p-\vec p')\,\delta_{NN'}\delta_{nn'}$,
\begin{equation}
 \begin{split}
 &(2\pi)^3\,M_{11}^{-9/2}\,\frac{R_-}{R_9^{5/2}}\,
 \frac{(n_1N_2-n_2N_1)^2N_3^2}{(N_1N_2)^2\sqrt{N_1N_2N_3}}\\
 &\quad\times(2\pi)^9\delta\bigl(\textstyle\sum E\bigr)\,\delta^8\bigl(\sum\vec p\bigr)\,
 \delta^8(\bar\eta_{\rm cm}+\bar\eta_3)\,\delta^8(\bar\eta_{\rm rel}),
 \end{split}
 \label{eq:HM}
\end{equation}
whose prefactor in our units is $(2\pi)^3M_{11}^{-9/2}R_-=(2\pi)^{3/2}$.

Our process of interest from Subsection~\ref{sec:amplitude} has $N_1=N_2=1$ and $N_3=2$, for which the $N$- and $n$-dependent factor in~\eqref{eq:HM} is $2\sqrt2\,(n_1-n_2)^2$. Real on-shell momenta, with energies $p^2/2N$ in our units for the nine-momentum $p=(\vec p,n/R_9)$, would have equal velocities $p_a/N_a$, where this factor vanishes, so~\cite{HerderscheeMaldacena} take the complex null momenta $p_a=(n_a/R_9)(\hat e_9+\ii\hat e_8)$ (their Eq.~(8)), of zero energy. As stated in the main text, we strip off the energy-conserving delta function and the $U(1)$ center-of-mass factor $(2\pi)^8\delta^8(\sum\vec p)\,\delta^8(\bar\eta_{\rm cm}+\bar\eta_3)$. The $\delta^8(\bar\eta_{\rm rel})$ means that only one unit-normalized state of the fermion zero modes $\hyperlink{def:coulomb}{\lambda_\alpha}$, of maximal helicity $\abs{h}=2$ in the 8--9 plane, has a nonvanishing amplitude, namely $\varepsilon_i\varepsilon_j\hyperlink{def:sij}{\ket{s_{ij}}}$, as in $\hyperlink{def:keps}{\ket{k,\varepsilon}}$ of~\eqref{eq:vertex}, with $\varepsilon=(\hat e_9+\ii\hat e_8)/\sqrt2\propto p_a$, so that $(n_1-n_2)^2=2R_9^2(q\cdot\varepsilon^*)^2$ for the relative momentum $q=(p_1-p_2)/2$.
The stripped amplitude then reads
\begin{equation}
 \begin{split}
 &(2\pi)^{3/2}R_9^{-5/2}\cdot2\sqrt2\cdot2R_9^2(q\cdot\varepsilon^*)^2\\
 &\quad=4\sqrt2\,(2\pi)^{3/2}R_9^{-1/2}\,(q\cdot\varepsilon^*)^2 .
 \end{split}
\label{toCancel}
\end{equation}

In \cite{HerderscheeMaldacena}, the $x^9$-direction was compactified in order to relate the amplitude to an index, with the result holding at any $R_9$. The matrix quantum mechanics is recovered as $R_9\to\infty$ at fixed $p_9=n/R_9$. In this limit each of the three external states picks up $\sqrt{2\pi R_9}$, so that $\braket{p}{p'}=(2\pi)^9\delta^9(p-p')$, while the Kronecker delta of the $n_a$ becomes $2\pi\delta(\sum p_9)/(2\pi R_9)$, whose momentum-conserving delta function we strip as well. Together these give $(2\pi R_9)^{3/2}/(2\pi R_9)=\sqrt{2\pi R_9}$, which cancels with the $R_9^{-1/2}$ of (\ref{toCancel}), giving the amplitude
\begin{equation}
 4\sqrt2\,(2\pi)^2\,(q\cdot\varepsilon^*)^2 .
 \label{eq:A-q}
\end{equation}

It remains to extract $\me{k,\varepsilon}{\hyperlink{def:keps}{H_{\rm int}}}{\psi_{\rm eff}}$ from~\eqref{eq:A-q}. Since on the \hyperlink{def:coulomb}{Coulomb branch} one can gauge-fix $\hyperlink{def:X}{\mathbf X_i}=r_i\sigma_3/\sqrt2$, the branes sit at $\pm\vec r/\sqrt2$, so the momentum conjugate to $\vec r$ is $k=\sqrt2\,q$, giving $(q\cdot\varepsilon^*)^2=\frac12(k\cdot\varepsilon^*)^2$. Since the center of mass is free, $\ket{p_3}=\ket{p_3}_{\rm cm}\ket{\psi_{\rm eff}}$, with $\ket{\psi_{\rm eff}}$ of~\eqref{eq:BO}, and $\ket{p_1}\ket{p_2}=2^{9/4}\ket{p_1+p_2}_{\rm cm}\ket{k,\varepsilon}$, with $\ket{p}_{\rm cm}$ the center-of-mass state of momentum $p$, normalized like $\ket p$, and $\braket{k}{k'}=(2\pi)^9\delta^9(k-k')$. The factor $2^{9/4}$ follows from $\delta^9(p_1-p_1')\,\delta^9(p_2-p_2')=2^{9/2}\delta^9(p_1+p_2-p_1'-p_2')\,\delta^9(k-k')$. The inner product of the center-of-mass states is the factor we stripped, so~\eqref{eq:A-q} is $2^{9/4}$ times our matrix element. Finally, the plane wave $\ket k$ is not a physical state: the gauge transformation exchanging the two identical branes maps $\vec r\to-\vec r$ and $\lambda_\alpha\to-\lambda_\alpha$~\cite{LinYin}, which leaves the states $\ket{s_{ij}}$ invariant, so in our channel gauge-invariant states are even under $\vec r\to-\vec r$. The physical two-brane state of relative momentum $k$, for which~\cite{HerderscheeMaldacena} compute the amplitude, is $(\ket k+\ket{-k})/\sqrt2$, with the normalization $(2\pi)^9[\delta^9(k-k')+\delta^9(k+k')]$, giving a factor of $\sqrt2$. Altogether,~\eqref{eq:A-q} becomes
\begin{equation}
 \begin{split}
 \me{k,\varepsilon}{H_{\rm int}}{\psi_{\rm eff}}&=2^{-1/2}\cdot2^{-9/4}\cdot4\sqrt2\,(2\pi)^2\cdot\tfrac12(k\cdot\varepsilon^*)^2\\
 &=2^{3/4}\pi^2(k\cdot\varepsilon^*)^2 ,
 \end{split}
\end{equation}
which is~\eqref{eq:vertex}.

\clearpage
\section{Invariant fermionic subspace}
\label{sec:fiber-proof}

The dimension and basis of the invariant subspace (denoted by $\mathcal{F}_*$ below) follow from
Schur--Weyl duality.
Recall the creation operators $f^\dagger_m=f^\dagger_{A\ell}$
of Appendix~\ref{sec:fermions}: $A$ is the fundamental index of
$\SU(4)\simeq\Spin(6)$, while $\ell=(c,\rho)=1,\ldots,6$ combines spherical
color and spin. The stabilizer acts on
$\ell$ through $K_4$ and on $A$ through $\SU(4)$, and fixes the vacuum.
Thus, writing $V=\C^6_{K_4}$ and $W=\C^4_{\SU(4)}$, we identify
\begin{equation}
 \mathcal F=\Lambda^\bullet(V\otimes W),\qquad
 \mathcal F_n=\Lambda^n(V\otimes W)
 =\bigoplus_{\lambda\vdash n}S_\lambda V\otimes S_{\lambda^{\mathsf T}}W.
 \label{eq:fermion-schur-weyl}
\end{equation}
Here $S_\lambda V$ and $S_{\lambda^{\mathsf T}}W$ are the irreducible
representations of the general linear groups $\operatorname{GL}(V)$
and $\operatorname{GL}(W)$ labeled by the Young diagram $\lambda$ and
its transpose $\lambda^{\mathsf T}$. The sum runs over diagrams with
at most six rows and four columns. This is the exterior-algebra form of
Schur--Weyl duality~\cite[Thm.~6.3 and Ex.~6.11(b)]{FultonHarris}.

The $\SU(4)$ invariants are precisely the summands with
$\lambda=(4^k)$, since only the four-row $\lambda^{\mathsf T}=(k,k,k,k)$,
built from fully contracted $\epsilon$ tensors, give $\SU(4)$ singlets.
Consequently they occur at fermion numbers $n=4k$, $0\leq k\leq6$,
and $\mathcal F_{4k}^{\SU(4)}\simeq S_{(4^k)}V$.
The remaining invariance is obtained by averaging over $K_4$ in each
summand:
\begin{equation}
 \Pi_k=\frac14\sum_{g\in K_4}S_{(4^k)}(g|_V),\qquad
 \mathcal F_*=\bigoplus_{k=0}^6\Pi_k S_{(4^k)}V.
 \label{eq:fiber-projector}
\end{equation}
Each of the three nonidentity elements acts on $V$ with eigenvalues
$\ii,\ii,\ii,-\ii,-\ii,-\ii$. Writing $s_\lambda$ for the Schur
polynomial, the dimension of each projected summand is therefore
\begin{equation}
 m_k=\Tr\Pi_k
 =\frac{s_{(4^k)}(1,1,1,1,1,1)
       +3s_{(4^k)}(1,1,1,-1,-1,-1)}4,
\end{equation}
where homogeneity of degree $4k$ removes the common factor $\ii$.
Evaluating the Schur polynomials gives
$(m_0,\ldots,m_6)=(1,36,468,1056,468,36,1)$ and hence
$\dim\mathcal F_*=\sum_km_k=2066$, as in~\eqref{eq:fiber-dimension}.

This derivation is constructive. For each $k$, take tableaux
$T=(q_{aj})$ of shape $(k,k,k,k)$ filled with
$\ell$-indices in $\{1,\ldots,6\}$, strictly increasing along rows and
weakly increasing down columns. Their transposes are the usual
semistandard tableaux of shape $(4^k)$. Map each column to creation
operators by
\begin{equation}
 \ket{B_T}=\prod_{j=1}^k
 B^\dagger_{q_{1j}q_{2j}q_{3j}q_{4j}}\ket0,\qquad
 B^\dagger_{\ell_1\ell_2\ell_3\ell_4}
 =\sum_{A_1,\ldots,A_4}\epsilon_{A_1A_2A_3A_4}
   \prod_{a=1}^4 f^\dagger_{A_a\ell_a},
 \label{eq:flavor-blocks}
\end{equation}
with $\epsilon_{1234}=1$. These states form the standard tableau basis
of $S_{(4^k)}V$;
their linear independence is part of the tableau construction.

To make the projection explicit, the half-turns
$\mathcal P_3=e^{\pi(\mathcal G_{12}+\mathcal R_{12})}$ and
$\mathcal P_1=e^{\pi(\mathcal G_{23}+\mathcal R_{23})}$ about the third and
first principal axes act,
according to~\eqref{eq:fermion-symmetry-action}, as
\begin{equation}
 \mathcal P_3f^\dagger_{A\ell}\mathcal P_3^{-1}
 =e^{-\ii\pi w_\ell}f^\dagger_{A\ell},\qquad
 \mathcal P_1f^\dagger_{A\ell}\mathcal P_1^{-1}
 =\ii f^\dagger_{A,7-\ell},\qquad
 w_\ell=c+\rho-\tfrac72.
 \label{eq:half-turn-modes}
\end{equation}
Set $w_T=\sum_{a,j}w_{q_{aj}}$ and
$(T^*)_{aj}=7-q_{5-a,k+1-j}$. The blocks are symmetric in their
$\ell$-indices and commute, so
$\mathcal P_3\ket{B_T}=(-1)^{w_T}\ket{B_T}$ and
$\mathcal P_1\ket{B_T}=\ket{B_{T^*}}$. Thus a basis of the projected
summand at fermion number $4k$ is
\begin{equation}
 \ket{e_T}=\Pi_k\ket{B_T}
 =\tfrac12\bigl(\ket{B_T}+\ket{B_{T^*}}\bigr),\qquad
 w_T\ \text{even},
 \label{eq:invariant-fiber-basis}
\end{equation}
with one representative of each orbit $\{T,T^*\}$.
Enumerating these states as $\ket{e_\mu}$ gives
\begin{equation}
 \mathcal F_*=\operatorname{span}\{\ket{e_\mu}:\mu=1,\ldots,2066\}.
 \label{eq:fiber-basis-enumerated}
\end{equation}

\clearpage
\section{Wick contractions and angular integrals}
\label{app:ansatz-wick}

The matrix elements used in Subsection~\ref{sec:ansatz} can be evaluated
by reducing to the gauge slice and performing Gaussian Wick contractions.
By~\eqref{eq:diagonal-wavefunction}, unitarity of
 $\mathcal U(O,V)$ gives
 \begin{equation}
  \braket{\psi_j(X)}{\psi_h(X)}
  =\braket{\psi_j(D(x))}{\psi_h(D(x))}.
 \end{equation}
 Integrating out $O,V$ gives the standard singular-value measure induced by
 $d^{27}X$~\cite[Eqs.~(3.9), (4.5)--(4.6)]{EdelmanRao}:
 \begin{equation}
  \begin{gathered}
   d\mu(x)=\frac{1024\pi^{13}}{4725}\prod_{r=1}^3 x_r^6
     \prod_{r<t}(x_r^2-x_t^2)\,d^3x,
     \qquad x_1>x_2>x_3>0,\\
   \int e^{-s}\,d\mu(x)=\pi^{27/2}.
  \end{gathered}
  \label{eq:ansatz-slice-measure}
 \end{equation}
 The radial moments and norm reduce to
 \begin{equation}
  \me{\psi_j}{R^{2k}}{\psi_h}
  =\int d\mu(x)\,s^k\braket{\psi_j(D(x))}{\psi_h(D(x))},
  \qquad k\geq0.
 \end{equation}
 The spinor covariance of $Q_\alpha$ similarly gives
 \begin{equation}
  \sum_{\alpha=1}^{16}
   \braket{(Q_\alpha\psi_j)(X)}{(Q_\alpha\psi_h)(X)}
  =\sum_{\alpha=1}^{16}
   \braket{(Q_\alpha\psi_j)(D(x))}{(Q_\alpha\psi_h)(D(x))}.
 \end{equation}
 The unitary fermion rotation and the spinor rotation cancel in this sum.
 The matrix elements of $H$ and $H_{\mathrm{aux}}$ therefore reduce to
 \begin{equation}
  \begin{aligned}
   \me{\psi_j}{H}{\psi_h}
    &=\frac1{16}\int d\mu(x)\sum_{\alpha=1}^{16}
      \braket{(Q_\alpha\psi_j)(D(x))}{(Q_\alpha\psi_h)(D(x))},\\
   \me{\psi_j}{H_{\mathrm{aux}}}{\psi_h}
    &=\frac1{16}\int d\mu(x)\,s\sum_{\alpha=1}^{16}
      \braket{(Q_\alpha\psi_j)(D(x))}{(Q_\alpha\psi_h)(D(x))}.
  \end{aligned}
 \end{equation}

 It therefore suffices to determine $\ket{(Q_\alpha\psi_j)(D(x))}$.
 Differentiating~\eqref{eq:diagonal-wavefunction} with respect to $X_{ai}$
 and evaluating at $X=D(x)$ gives, with $\partial_{ai}=\partial/\partial X_{ai}$,
 \begin{equation}
  \begin{aligned}
   \left.\partial_{aa}\ket{\psi_j(X)}\right|_{X=D(x)}
    &=\partial_{x_a}\ket{\psi_j(D(x))},\\
   \left.\partial_{ab}\ket{\psi_j(X)}\right|_{X=D(x)}
    &=\frac{x_b\mathcal G_{ab}+x_a\mathcal R_{ab}}{x_b^2-x_a^2}
       \ket{\psi_j(D(x))},\qquad a<b\leq3,\\
   \left.\partial_{ba}\ket{\psi_j(X)}\right|_{X=D(x)}
    &=\frac{x_a\mathcal G_{ab}+x_b\mathcal R_{ab}}{x_b^2-x_a^2}
       \ket{\psi_j(D(x))},\qquad a<b\leq3,\\
   \left.\partial_{ai}\ket{\psi_j(X)}\right|_{X=D(x)}
    &=-\frac{\mathcal R_{ai}}{x_a}\ket{\psi_j(D(x))},\qquad i>3.
  \end{aligned}
  \label{eq:ansatz-slice-derivatives}
 \end{equation}
 Substituting these identities in~\eqref{eq:supercharge} gives
 $\ket{(Q_\alpha\psi_j)(D(x))}$ from the wavefunction on the gauge slice
 and its three derivatives.

For the profiles in~\eqref{eq:invariant-profiles}, the diagonal
 derivatives in~\eqref{eq:ansatz-slice-derivatives} give
 \begin{equation}
  \partial_{aa}f_j
  =-\frac{x_ae^{-a_js/2}}{4\pi}\int_{S^2}\!d\Omega(n)\,
    \bigl[a_j+(b_j-a_j)n_a^2\bigr]
    e^{(a_j-b_j)\sum_r n_r^2x_r^2/2}.
 \end{equation}
 All remaining derivatives of $f_j$ vanish on the gauge slice by invariance.
 Acting on the
 polynomial $\ket{P_j(X)}$, the $\partial_{ai}$ again give polynomials
 in $X$, so the denominators in~\eqref{eq:ansatz-slice-derivatives} cancel.

 Thus, after extracting the two oriented Gaussians, the derivatives and
 fermionic contractions in the matrix elements above leave a scalar
 polynomial $p(D(x)D(x)^{\mathsf T};n,m)$ on the gauge slice.
 The product of the two Gaussians has weight
 \begin{equation}
  e^{-\tr(KD(x)D(x)^{\mathsf T})},\qquad
  K=\frac{a_j+a_h}{2}\one
    -\frac{a_j-b_j}{2}nn^{\mathsf T}
    -\frac{a_h-b_h}{2}mm^{\mathsf T}.
 \end{equation}
 The integrand is invariant under a simultaneous rotation of $D(x)$, $n$, and
 $m$ by a color matrix. Thus, after averaging over $n,m$, we can replace the
 integral over the gauge slice by a Gaussian integral over all entries of $X$:
 \begin{equation}
  \begin{aligned}
   &\frac1{(4\pi)^2}\int d\mu(x)\,
    \int_{S^2}\!d\Omega(n)\int_{S^2}\!d\Omega(m)\,
    e^{-\tr(KD(x)D(x)^{\mathsf T})}p(D(x)D(x)^{\mathsf T};n,m)\\
   &\qquad=\frac{\pi^{27/2}}{(4\pi)^2}
    \int_{S^2}\!d\Omega(n)\int_{S^2}\!d\Omega(m)\,
    \frac{\langle p(XX^{\mathsf T};n,m)\rangle_{X,K}}{(\det K)^{9/2}},
  \end{aligned}
  \label{eq:ansatz-profile-moments}
 \end{equation}
 where
 \begin{equation}
  \langle\cdots\rangle_{X,K}
  :=\frac{(\det K)^{9/2}}{\pi^{27/2}}
    \int_{\mathbb R^{3\times9}}\!d^{27}X\,
    e^{-\tr(KXX^{\mathsf T})}\,\cdots\,.
 \end{equation}
 Wick contraction evaluates $\langle p(XX^{\mathsf T};n,m)\rangle_{X,K}$, leaving only
 the two angular integrals
 over $n$ and $m$ in~\eqref{eq:ansatz-profile-moments}.

To evaluate these remaining integrals, set
 \begin{equation}
  \gamma=\frac{a_j+a_h}{2},\qquad
  \eta_j=\frac{a_j-b_j}{2},\qquad
  \eta_h=\frac{a_h-b_h}{2},\qquad z=n\cdot m,
 \end{equation}
 \begin{equation}
  \Delta(z)=\Delta_0-\eta_j\eta_h z^2,\qquad
  \Delta_0=(\gamma-\eta_j)(\gamma-\eta_h),\qquad
  \Delta_1=\Delta(1)=\gamma\frac{b_j+b_h}{2}.
 \end{equation}
 Then $K=\gamma\one-\eta_jnn^{\mathsf T}-\eta_hmm^{\mathsf T}$ and
 $\det K=\gamma\Delta(z)$, with
 \begin{equation}
  (K^{-1})_{ab}=\frac{\delta_{ab}}{\gamma}
  +\frac{\eta_j(\gamma-\eta_h)n_an_b
         +\eta_h(\gamma-\eta_j)m_am_b
         +\eta_j\eta_h z(n_am_b+m_an_b)}{\gamma\Delta(z)}.
 \end{equation}
 The contracted moments are rational functions of $z$;
 the integrand is invariant under simultaneous color rotations and under
 $n\mapsto-n$, $m\mapsto-m$, so
 \begin{equation}
  \frac1{(4\pi)^2}\int_{S^2}\!d\Omega(n)\int_{S^2}\!d\Omega(m)
  =\frac1{8\pi}\int_{S^2}\!d\Omega(n)\int_{-1}^1dz
  =\frac12\int_{-1}^1dz
  =\int_0^1dz.
 \end{equation}
 For integers $L\geq0$ and $0\leq r\leq L+1$, the substitution
 $y=z/\sqrt{\Delta(z)}$ gives the required integrals:
 \begin{equation}
  \int_0^1\frac{z^{2r}\,dz}{\Delta(z)^{9/2+L}}
  =\Delta_0^{r-4-L}
   \sum_{t=0}^{3+L-r}\binom{3+L-r}{t}
   \frac{(\eta_j\eta_h)^t}
   {(2r+2t+1)\Delta_1^{r+t+1/2}}.
  \label{eq:ansatz-orientation-sum}
 \end{equation}
Together with the Wick contractions, this finite sum gives the norm,
radial moments, energy, and auxiliary-Hamiltonian matrix elements.

\end{document}